\documentclass [11pt, a4paper, notitlepage] {article}

\usepackage [english] {babel}
\usepackage [cp1250] {inputenc}
\usepackage [T1] {fontenc}
\usepackage {graphicx, epsfig}
\usepackage {amsmath}
\usepackage {amssymb}
\usepackage[dvips]{color}
\usepackage{array}
\usepackage{hyperref}

\begin{document}

\bibliographystyle{plane}

\title{Embedding $Z'$ models in SO(10) GUTs}
\date{Received: 5.12.2012 / Revised version: 23.04.2013}
\author{Pawe{\l} Pacho{\l}ek \\
Institute of Theoretical Physics, Faculty of Physics, \\
University of Warsaw, Ho{\.z}a 69, Warsaw, Poland \\
\textit{Pawel.Pacholek@fuw.edu.pl}}

\definecolor{brown}{rgb}{0.6,0.2,0.1}
\definecolor{violet}{rgb}{0.40,0,0.70}

\maketitle

\abstract{We embed a theory with $Z'$ gauge boson (related to extra $U(1)$ gauge group) into a supersymmetric GUT theory based on $SO(10)$. Two possible sequences of $SO(10)$ breaking via VEVs of appropriate Higgs fields are considered. Gauge coupling unification provides constraints on low energy values of two additional gauge coupling constants related to $Z'$ interactions with fermions. Our main purpose is to investigate in detail the freedom in these two values due to different scales of subsequent $SO(10)$ breaking and unknown threshold mass corrections in the gauge RGEs. These corrections are mainly generated by Higgs representations and can be large because of the large dimensions of these representations. To account for many free mass parameters, effective threshold mass corrections have been introduced. Analytic results that show the allowed regions of values of two additional gauge coupling constants have been derived at 1-loop level.  For a few points in parameter-space that belong to one of these allowed regions 1-loop running of gauge coupling constants has been compared with more precise running, which is 2-loop for gauge coupling constants and 1-loop for Yukawa coupling constants. 1-loop results have been compared with experimental constraints from electroweak precision tests and from the most recent LHC data.
%
\vspace{0.5cm}\newline Journal Keywords: $Z'$, GUT, unification, gauge couplings, RGE, B-L, extra U(1), threshold corrections, Yukawa couplings, Planck Mass
}


\section{Introduction}{\label{SecInt}}

Models with additional $Z'$ gauge boson are simple extensions of the Standard Model and have been extensively studied in the literature \cite{Aguila}, \cite{Aguila2}, \cite{Leike}, \cite{Perez}, \cite{Perez2}, \cite{Appelquist}, \cite{Lopez}, \cite{Zwirner1}, \cite{Zwirner2}, \cite{Erler}, \cite{Kim}. They are based on the gauge group, which is $SU(3)_{c} \oplus SU(2)_{L} \oplus U(1)^{2}$ and will be called the $Z'$ gauge group or denoted shortly by $G_{3211}$. Special attention has been paid to models, where the $U(1)^{2}$ algebra is spanned by the weak hypercharge $Y$ and the $B-L$ quantum number. Thus, it's the only case where the theory, to be anomaly free, doesn't require additional, exotic fermions and the fermionic spectrum of the Standard Model (SM) has to be supplemented only by three right-handed neutrinos.

Such models have 3 additional to the SM free parameters, two abelian gauge coupling constants and the $Z'$ mass - $M_{Z'}$. One of the important points of the LHC experimental program is the search for a $Z'$ boson and, indeed, interesting limits on $M_{Z'}$ and its couplings are already available and will be gradually improved \cite{ATLAS},\cite{CMS}.
It becomes then possible to ask if the experimental limits on the parameters of the additional $Z'$ (or in case of its discovery - the values of these parameters) are consistent with an UV completion of such effective low energy model by its embedding into a GUT theory. This question has been studied in a number of papers \cite{Lopez}, \cite{Zwirner1}, \cite{Zwirner2}, \cite{Erler}, \cite{Kim}.

In the present paper we readdress this question in the case of the supersymmetric $SO(10)$ GUT theory. $SO(10)$ is one of the simplest possible GUT gauge groups in which $G_{3211}$ can be embedded. Our main purpose is to investigate in detail the freedom in the values of the low energy gauge couplings of the unified theory due to different scales of subsequent $SO(10)$ breaking and unknown threshold mass corrections in the gauge RGEs. These corrections are mainly generated by Higgs representations and can be large because of the large dimensions of these representations, which are necessary to break $SO(10)$ to $G_{3211}$ and further to the SM gauge group. Our approach is conservative in the sense of allowing for large mass splittings in the spectrum. We also consider the influence of 2-loop effects on gauge couplings.

In section \ref{SecU1N} we recall the formalism of the theories with additional $U(1)$ factors. This formalism, in the case of only one such factor, is then adopted to describe the $Z'$ gauge theory originating from the $SO(10)$-breaking. In section \ref{SecSO10} we discuss various possible patterns of $SO(10)$ breaking and the Higgs representations that are needed to realize them.  Then we introduce two specific models for further, detailed analysis. No attempt is made to present the dynamical theory of breaking, but merely the group theoretical aspects of the necessary spectra are given. The region of possible values of gauge coupling constants in these models, enlarged by unknown threshold contributions from Higgs multiplets and other fields, is given in section \ref{Sec1loop}. Effective threshold corrections are introduced there to account for many free mass parameters and analytical results are presented. Two-loop effects are discussed in section \ref{Sec1,5loop}. A comparison with experimental constraints is presented in section \ref{SecDosw}. We conclude in section \ref{SecPods}.

\section{\texorpdfstring{Models with $U(1)^{N}$ gauge group - formalism and\newline parametrization}{\space}}{\label{SecU1N}}

The most useful form of the Lagrangian for a theory with unbroken $U(1)^{N}$ gauge symmetry contains the following, abelian part

\begin{equation}\begin{array}{l}
L^{abelian} = -\frac{1}{4}F_{a}^{\mu \nu} F_{a \mu \nu} +\\ + i\sum_{f} \bar{\Psi}_{f}\gamma_{\mu}\left(\partial^{\mu} -i (X^{f\,T})_{a} G_{a b} A^{\mu}_{b} \right) \Psi_{f} \\ + \sum_{s} \left[\partial^{\mu}\varphi_{s} - i \left((X^{s\,T})_{a} G_{a b} A^{\mu}_{b} \right) \varphi_{s} \right]\\ \left[\partial_{\mu}\varphi_{s}^{*} + i \left((X^{s\,T})_{c} G_{c d} A_{\mu d} \right)\varphi_{s}^{*} \right]\label{L2}\end{array}
\end{equation}
Lower case Latin index $s$ denotes a single complex scalar field. Analogously, lower case Latin index $f$ denotes a single fermionic (Weyl) field. $\Psi_{f}$ is a chiral field (left-handed or right-handed) in Dirac notation. Lower case Latin indices $a,b,c,d \in \left\{1,2,\ldots..N\right\}$ are indices of abelian $U(1)$ gauge groups. $A_{a}^{\mu}$ denotes the abelian gauge field, $F_{a}^{\mu \nu} = \partial^{\mu} A_{a}^{\nu} - \partial^{\nu} A_{a}^{\mu}$, $G_{ab}$ is the matrix of abelian gauge coupling constants, $(X^{x})_{a}$ is the charge of the single field $x$ ($s$ or $f$) with respect to the abelian generator $X_{a}$.

The form of the Lagrangian given in eq. (\ref{L2}) is not the most general one, because the most general abelian kinetic term is $-\frac{1}{4}F_{a}^{\mu \nu} \widetilde{h}_{ab} F_{b \mu \nu}$. However, one can use linear transformations of fields $A_{a}^{\nu}$ to diagonalize this term and constrain oneself to the case, where $\widetilde{h}_{ab} = \delta_{ab}$ at tree level. This equality is not affected by RGE running if the relation between bare and renormalized kinetic terms is the following:
\begin{equation}
-\frac{1}{4}F_{a}^{bare\,\mu \nu} F^{bare}_{a \mu \nu} = -\frac{1}{4}\left(Z_{3\,ab} - \delta_{ab}\right) F_{a}^{\mu \nu} F_{b \mu \nu} -\frac{1}{4} F_{a}^{\mu \nu} F_{a \mu \nu} 
\end{equation}
where $Z_{3\,ab}$ is the matrix of counterterms. It should\newline usually be nondiagonal to cancel nondiagonal divergences that appear in the abelian self-energy Green function at one-loop. This Green function can also have finite nondiagonal terms, but at the tree-level (which is just the abelian propagator) it's always diagonal.

There is a large freedom of transformations, that preserve the form of the Lagrangian given in eq. (\ref{L2}). Firstly, one can transform fields $A_{a}^{\mu}$ orthogonally

\begin{equation}A^{\nu}_{a'} = O_{a' b} A^{\nu}_{b}\label{Otrans}\end{equation}
where $O_{a' b}$ is an orthogonal matrix. Secondly, one can transform abelian generators $X_{a}$ linearly

\begin{equation} X_{a'} = L_{a' b} X_{b}\label{Ltrans}\end{equation}
where $L_{a' b}$ is a linear, invertible matrix. Gauge coupling matrix $G_{a b}$ transforms under eq. (\ref{Otrans}) and (\ref{Ltrans}) in the following way

\begin{equation}
G_{a' b'} = (L^{T})^{-1}_{a' b} G_{b a} O^{T}_{a b'}
\label{Gprim}
\end{equation}
For any given basis of abelian generators $X_{a}$ one can always use $O_{a' b}$ transformation to make $G_{a b}$ matrix upper-triangular with non-negative diagonal terms, so we have only $N(N+1)/2$ physical abelian gauge coupling constants \cite{Aguila}. However, this upper-triangularity is generally not preserved under RGE running, so it seems that one still has to solve RGE system with $N^{2}$ independent abelian coupling constants. To get rid of unphysical degrees of freedom from the RGE system one can introduce $O$-invariant symmetric product $\vartheta_{ab}=\frac{1}{4\pi}(GG^{T})_{ab}$ which is uniquely related to the upper-triangular form of $G_{a b}$ matrix (in particular it has the same number of independent parameters). The RGE for $\vartheta_{ab}$ are given in Appendix \ref{ApRGE}. For $N=2$, assuming $G_{21}=0$, we have

\begin{equation}
\vartheta_{ab} = \frac{1}{4\pi} \left[\begin{array}{cc}
G_{11}^{2} + G_{12}^{2} & G_{12}G_{22} \\
     G_{12}G_{22}       & G_{22}^{2}  
\end{array}\right]
\label{Gu-t2}
\end{equation}
Inverting the above relation leads to

\begin{equation}\begin{array}{l}
G_{22} = 2\sqrt{\pi}\cdot\sqrt{\vartheta_{22}} \hspace{1cm} G_{12} = 2\sqrt{\pi}\cdot\frac{\vartheta_{12}}{\sqrt{\vartheta_{22}}} \\ G_{11} = 2\sqrt{\pi}\cdot \sqrt{\vartheta_{11}-\frac{\vartheta_{12}^{2}}{\vartheta_{22}}}
\end{array}\end{equation}

The $\vartheta_{ab}$ matrix is $O$-invariant, but not $L$-invariant and it transforms under eq. (\ref{Gprim}) in the following way:

\begin{equation}
\vartheta_{a' b'} = (L^{T})^{-1}_{a' b} \vartheta_{b a} L^{-1}_{a b'}
\label{Aprim}
\end{equation}
Transformations $L_{a' b}$ are useful, when there are at least two different bases of abelian generators $X_{a}$, which are important due to some specific properties. Such a situation is present, when the theory with $U(1)^{N}$ gauge symmetry is a low-energy effective theory coming from GUTs. Then, there is a special basis of $X_{a}$ generators, in which the unification is explicit. Generators of this basis will be denoted by $X_{\widehat{a}}$ and they are standard, diagonal generators of the GUT group. Such a special basis is of course natural at high energy, near the GUT scale. However, at lower scale the $U(1)^{N}$ gauge symmetry should be broken down to $U(1)_{Y}$ (corresponding to weak hypercharge $Y$). Therefore, at low scale it's natural to take $Y$ as one of the basis-generators and complete the basis with generators related to other quantum numbers, that are important at relatively low energy. These generators will be denoted by $X_{\underline{a}}$. The natural low-energy basis obtained this way is usually different than $X_{\widehat{a}}$ basis.

We shall consider in this paper the low energy effective theory with additional (with respect to the SM gauge symmetry) $U(1)$ gauge symmetry and with the fermion spectrum of the SM, supplemented by three right-handed neutrinos. The gauge group of such a theory is $G_{SM} \oplus U(1)$, where $G_{SM}$ is the SM gauge group. It's well known, that with such a fermion spectrum the only anomaly-free $U(1)$ gauge groups are $U(1)_{Y}$, $U(1)_{B-L}$ and their linear combinations \cite{Appelquist}. Therefore, the natural low energy basis is $(Y, B-L)$. The purpose of this paper is to investigate constraints on that low energy effective theory, once it is embedded into $SO(10)$ GUT theory. Those constraints depend on the assumed breaking pattern of $SO(10)$. In our conservative approach there is an intermediate symmetry breaking scale and constraints are weaker than in the one step symmetry breaking. Also a convenient choice of the $X_{\widehat{a}}$ basis $(X_{\widehat{1}},X_{\widehat{2}})$ depends on the breaking pattern of $SO(10)$. When $SO(10)$ is initially broken to $SU(5) \oplus U(1)_{X}$ and SM-fermions are embedded into three $16$'s of $SO(10)$, a natural choice of $X_{\widehat{a}}$ is such, that

\begin{equation} \text{Pattern I:}\;
\left[\begin{array}{c}
X_{\widehat{1}} \\
X_{\widehat{2}}
\end{array}\right] = \left[\begin{array}{c}
\widehat{Y} \\
X
\end{array}\right] = \left[\begin{array}{cc}
\frac{\sqrt{15}}{5} & 0 \\
-\frac{\sqrt{10}}{5} & \frac{\sqrt{10}}{4}
\end{array}\right] \cdot \left[\begin{array}{c}
Y \\
B-L
\end{array}\right]\label{case1}\end{equation}
As we can see, the weak hypercharge $Y$ is rescaled to $\widehat{Y}$ just like in the minimal GUT model, based on $SU(5)$. When $SO(10)$ is initially broken to $SU(3) \oplus SU(2)_{L} \oplus SU(2)_{R} \oplus U(1)_{B-L}$, we take as $(X_{\widehat{1}},X_{\widehat{2}})$ basis

\begin{equation} \text{Pattern II:}\;
\left[\begin{array}{c}
X_{\widehat{1}} \\
X_{\widehat{2}}
\end{array}\right] = \left[\begin{array}{c}
R\\
\widehat{B-L}
\end{array}\right] = \left[\begin{array}{cc}
1 & -\frac{1}{2} \\
0 & \frac{\sqrt{6}}{4}
\end{array}\right] \cdot \left[\begin{array}{c}
Y \\
B-L
\end{array}\right]\label{case2}\end{equation}
$R$ is the third (diagonal) generator of $SU(2)_{R}$ and $\widehat{B-L}$ is the appropriately rescaled $B-L$.
Formulas (\ref{case1}) and (\ref{case2}) are examples of formula (\ref{Ltrans}). There are also other breaking patterns of $SO(10)$ (with other intermediate groups), which may lead to $G_{3211}$. However, for all of them the $(X_{\widehat{1}},X_{\widehat{2}})$ basis is the same as in Pattern II.

$L_{a' b}$ transformations can in general be scale-dependent which would lead to running charges (eigenvalues of generators). However, it's much easier to solve the RGE system if charges are scale-independent. In order to do it, one has to write the RGE system in the concrete basis of abelian generators $X_{a}$. The basis of $X_{\widehat{a}}$ generators and the low-energy basis are obviously two most natural and useful ones for solving the RGE system.

In the basis of $X_{\widehat{a}}$ generators, at the GUT-breaking scale, one can use $O_{a' b}$ transformation to make the $G_{\widehat{a} \widehat{b}}$ matrix not only upper-triangular but even diagonal and then its diagonal terms should be unified with appropriate non-abelian gauge coupling constants $g_{A}$, according to the specific pattern of GUT-breaking (index $A$ denotes a simple subgroup of the total gauge group below GUT-breaking scale). Therefore, the $\vartheta_{\widehat{a} \widehat{b}}$ matrix is also diagonal and its diagonal terms should be unified with appropriate non-abelian gauge coupling constants $\alpha_{A}=\frac{1}{4\pi}g_{A}^{2}$.

For instance, for Pattern I, the $SO(10)$ group is broken to $SU(5) \oplus U(1)_{X}$ with gauge coupling constants $g_{5}$ and $g_{X}$ respectively. At the $SO(10)$ breaking scale denoted by $\mu_{0}$ these gauge coupling constants are unified
\begin{equation}g_{5}(\mu_{0})  = g_{X}(\mu_{0}) = g_{10}(\mu_{0}) \label{unif1h}\end{equation}
where $g_{10}$ is the unified gauge coupling constant of $SO(10)$ gauge group. RG evolution below the scale $\mu_{0}$ splits the coupling constants $g_{5}(\mu) \neq g_{X}(\mu)$ for $\mu_{1} < \mu < \mu_{0}$, where the scale $\mu_{1}$ is the scale of another symmetry breaking $SU(5) \oplus U(1)_{X} \rightarrow SU(3)_{c} \oplus SU(2)_{L} \oplus U(1)^{2}$. At this scale, the Lagrangian contains the following part

\begin{equation}\begin{array}{l}
L^{gauge}_{2\,int} \supset \sum_{f} \bar{\Psi}_{f}\gamma_{\mu} \left(X^{f} g_{X} A_{X}^{\mu} + \widehat{Y}^{f} g_{\widehat{Y}} A_{\widehat{Y}}^{\mu} +\right.\\ \left.+ \sum_{\alpha=1}^{3} T^{f}_{\alpha} g_{2} W_{\alpha}^{\mu} + \sum_{\alpha=1}^{8} J^{f}_{\alpha} g_{3} g_{\alpha}^{\mu}\right)\Psi_{f} \end{array}
\label{L2int}
\end{equation}
and the diagonal ($\widetilde{h}_{ab} = \delta_{ab}$) gauge kinetic term. The $G_{a b}$ matrix is also diagonal

\begin{equation}
G_{a b}= G_{\widehat{a} \widehat{b}} = \left[\begin{array}{cc}
g_{\widehat{Y}} & 0 \\
0 & g_{X}
\end{array}\right]
\label{Gin1}
\end{equation}
Below the scale $\mu_{1}$ the coupling constants split further

\begin{equation}g_{2}(\mu) \neq g_{3}(\mu) \neq g_{\widehat{Y}}(\mu) \label{unif1l}\end{equation}
Moreover, the matrix $G_{\widehat{a} \widehat{b}}$ (and $\vartheta_{\widehat{a} \widehat{b}}$) is no longer diagonal because of the RGE-running which generates a nondiagonal abelian gauge coupling constant.

In the above example two $U(1)$ groups are not necessarily unified directly with each other. $U(1)_{\widehat{Y}}$ is unified with $SU(3)_{c}$ and $SU(2)_{L}$ to $SU(5)$ at the scale $\mu_{1}$ and $U(1)_{X}$ could be unified with $SU(5)$ to $SO(10)$ at higher scale $\mu_{0}$.
In the special case of GUT-breaking when $\mu_{0}=\mu_{1}$, $U(1)$ groups are unified directly with each other and matrices $G_{\widehat{a} \widehat{b}}$ and $\vartheta_{\widehat{a} \widehat{b}}$ are not only diagonal but even proportional to identity \cite{Aguila}. 

Summarizing, unification is a source of constraints on the space of parameters (especially gauge coupling constants) and these constraints are most obvious and simple in the basis of $X_{\widehat{a}}$ generators, at the GUT-breaking scale. We are interested in expressing these constraints at low energy-scale (which is available in LHC) in the low-energy basis which is natural at this scale. Therefore, we need to calculate the RGE-running and apply the appropriate $L_{\widehat{a} \underline{b}}$ transformation. As it was already mentioned, constraints on the low energy theory, coming from embedding it into $SO(10)$ GUT theory, depend on the assumed pattern of the $SO(10)$ breaking and the necessary Higgs field spectrum. The latter is relevant, as we want to investigate the effect of the high mass threshold corrections in the gauge RGE running.

\section{Relating \texorpdfstring{$Z'$}{Z'} models to \texorpdfstring{$SO(10)$}{SO(10)} GUTs}{\label{SecSO10}}

Any possible breaking of $SO(10)$ down to\newline $SU(3)_{c} \oplus SU(2)_{L} \oplus U(1)^{2}$ has to be rank-conserving. Such a breaking requires appropriate representations of higgs fields which have to contain zero-weights. There are only 3 $SO(10)$-representations that contain zero-weights and have dimensions less than 600: $45$, $54$ and $210$ (they are all self-conjugated)\cite{slansky}. In order to avoid using really huge representations, one has to use only some of these 3 ones to realize the desired breaking. Moreover, $SU(3)_{c} \oplus SU(2)_{L} \oplus U(1)^{2}$ group is not a maximal little group of any of those 3 representations. Therefore, a two-step symmetry breaking with one intermediate scale can be realized.
We consider a class of theories with two such higgs representations that may obtain two significantly different VEVs related to two different energy scales. As in the previous section, the higher scale, related directly to $SO(10)$ breaking, will be denoted by $\mu_{0}$ and the other one will be denoted by $\mu_{1}$. Of course, the degenerate case ($\mu_{0} = \mu_{1}$), which has been considered in ref. \cite{Zwirner1}, \cite{Zwirner2}, is not excluded and will be also taken into account. We assume that $\mu_{0} \leq M_{Pl}$.

The $Z'$ boson can be light enough for possible detection in the LHC only if the $SU(3)_{c} \oplus SU(2)_{L} \oplus U(1)^{2}$ symmetry is broken at a relatively low energy scale (of the order of at most few TeV) to the SM gauge group. This breaking scale will be denoted by $\mu_{2}$. It requires a higgs field which has to be a singlet under SM gauge group but has to be charged under $U(1)^{2}$ gauge group. It will be denoted by $\chi$. To have complete theory one has to embed $\chi$ into appropriate representation of $SO(10)$. Taking into account only representations with dimensions less than 600, one has the following possibilities: $16$, $\overline{16}$, $126$, $\overline{126}$, $144$, $\overline{144}$,  $560$ and $\overline{560}$ \cite{slansky}.

Finally, one has to properly embed standard Higgs doublets $H_{u}$ and $H_{d}$. There are the following possibilities: $10$, $120$, $126$, $\overline{126}$, $210'$ and $320$ \cite{slansky}. Another thing that must be done is embedding SM-fermions and their Yukawa couplings. The minimal choice is to embed each generation in one $16$ of $SO(10)$. Then the embedding of SM Yukawa couplings into $SO(10)$-invariant structures requires including at least one of three $SO(10)$ higgs - containing irreps: $10$, $120$ or $\overline{126}$. Each of them, when coupled to two $16$-s, can produce a singlet that becomes the unified Yukawa coupling (because $16\otimes 16 = 10 \oplus 120 \oplus 126$ and $10$,$120$ are self-conjugated). $126$, $210'$, $320$ or any other irrep of $SO(10)$ doesn't have this property. Therefore, the general, unified form of these Yukawa couplings (which will be called $16$-Yukawa couplings) is the following:

\begin{equation}\frac{1}{2}\,16_{i}\,(Y_{10}^{ij}10+Y_{120}^{ij}120+Y_{\overline{126}}^{ij}\overline{126})\,16_{j}\end{equation} 
$Y_{10}^{ij}$, $Y_{120}^{ij}$ and $Y_{\overline{126}}^{ij}$ are unified Yukawa coupling constants. $Y_{120}^{ij}$ is anti-symmetric in generational indices $i$ and $j$ while $Y_{10}^{ij}$ and $Y_{\overline{126}}^{ij}$ are symmetric. According to \cite{Joshipura} and \cite{Bertolini}, these\newline Yukawa couplings can generate the correct spectrum of the SM and properties of neutrinos without additional terms that break $SO(10)$ explicitly. It's possible due to additional Higgs doublets that are embedded in both $120$ and $\overline{126}$ of $SO(10)$. As Higgs doublets embedded in $10$ of $SO(10)$, they may obtain VEVs which generate masses of SM fermions. Larger number of such Higgs doublets means larger number of free parameters that can be fitted to the SM spectrum.

Considerations included in the above part of this section can be summarized in one table (Table \ref{tab1}) that contains possible symmetry breaking scenarios for $Z'$ models embedded in $SO(10)$ GUTs.
\begin{table*}
\caption{Symmetry breaking scenarios for $SO(10)\rightarrow SU(3)_{c} \oplus SU(2)_{L} \oplus U(1)^{2}$. Dimensions of Higgs representations with respect to simple non-abelian gauge groups are written in square brackets. The order of these dimensions is always the same as the order of corresponding groups in the name of the whole gauge group. $U(1)$ charges have been omitted, but they are uniquely determined\cite{slansky}. Right lower index is a dimension of an $SO(10)$ representation in which a given Higgs representation is embedded. Separating higgs representations with commas means that each of them can provide a given symmetry breaking alone, without other ones. $X_{\widehat{a}}$ generators are defined in eq. (\ref{case1}) and (\ref{case2}).}
\begin{tabular}{||c||c|c|c|c||c||}
\hline \hline Names        & \multicolumn{4}{c||}{Groups and possible higgses}                                                 & Scales    \\
\hline\hline Initial group & \multicolumn{4}{c||}{$\textcolor{violet}{SO(10)}$}                                                &  $\textcolor{violet}{\mu_{0}} - M_{Pl}$ \\
\hline Breaking & I & II & III & IV &    \\
 patterns &  &  &  &  & $\textcolor{violet}{\mu_{0}}$ \\
\cline{1-5} Possible higgses & $[\textcolor{violet}{45}]$, $[\textcolor{violet}{210}]$           & $[\textcolor{violet}{45}]$                   & $[\textcolor{violet}{45}]$              & $[\textcolor{violet}{54}]$, $[\textcolor{violet}{210}]$        & \\
\hline\hline Possible         &                                 & $SU(3)_{c}\oplus$                   & $\textcolor{blue}{SU(4)}\oplus$       & $\textcolor{blue}{SU(4)}\oplus$ & \\
       intermediate     & $\textcolor{blue}{SU(5)}\oplus$ & $SU(2)_{L}\oplus$                   & $SU(2)_{L}\oplus$ & $SU(2)_{L}\oplus$ & $\textcolor{blue}{\mu_{1}} - \textcolor{violet}{\mu_{0}}$          \\
       (maximal little) & $U(1)_{X}$                      & $\textcolor{blue}{SU(2)_{R}}\oplus$ & $U(1)_{R}$         & $\textcolor{blue}{SU(2)_{R}}$        &  \\
       groups           &                                        & $U(1)_{B-L}$                        &          &            &           \\       
\hline Possible         & $[\textcolor{blue}{24}]_{\textcolor{blue}{45}}$,$[\textcolor{blue}{24}]_{\textcolor{blue}{54}}$,  & $[1,1,\textcolor{blue}{3}]_{\textcolor{blue}{45}}$,        & $[\textcolor{blue}{15},1]_{\textcolor{blue}{45}}$,     & $[\textcolor{blue}{15},1,\textcolor{blue}{3}]_{\textcolor{blue}{210}}$,   &  \\
       higgses          & $[\textcolor{blue}{24}]_{\textcolor{blue}{210}}$, & $[1,1,\textcolor{blue}{3}]_{\textcolor{blue}{210}}$        & $[\textcolor{blue}{15},1]_{\textcolor{blue}{210}}$    &  $\left([\textcolor{blue}{15},1,1]+\right.$                  &   $\textcolor{blue}{\mu_{1}}$        \\
                 & $[\textcolor{blue}{75}]_{\textcolor{blue}{210}}$ &         &     &  $\left.[1,1,\textcolor{blue}{3}]\right)_{\textcolor{blue}{45}}$                  &          \\       
\cline{1-5} $X_{\widehat{a}}$ generators & $\widehat{Y}, X$ & \multicolumn{3}{c||}{$R, \widehat{B-L}$} &    \\   
\hline\hline $G_{3211}$         & \multicolumn{4}{c||}{$SU(3)_{c} \oplus SU(2)_{L} \oplus \textcolor{red}{U(1)^{2}}$} & $\textcolor{red}{\mu_{2}} - \textcolor{blue}{\mu_{1}}$ \\
\hline Possible & \multicolumn{4}{c||}{$[1,1]_{\textcolor{red}{16}}$,$[1,1]_{\textcolor{red}{\overline{16}}}$,$[1,1]_{\textcolor{red}{126}}$,$[1,1]_{\textcolor{red}{\overline{126}}}$} &  \\
 higgses & \multicolumn{4}{c||}{$[1,1]_{\textcolor{red}{144}}$,$[1,1]_{\textcolor{red}{\overline{144}}}$,$[1,1]_{\textcolor{red}{560}}$,$[1,1]_{\textcolor{red}{\overline{560}}}$} & $\textcolor{red}{\mu_{2}}$ \\
\hline\hline SM group         & \multicolumn{4}{c||}{$SU(3)_{c} \oplus \textcolor{brown}{SU(2)_{L}} \oplus \textcolor{brown}{U(1)_{Y}}$} & $\textcolor{brown}{M_{Z}} - \textcolor{red}{\mu_{2}}$\\
\hline Possible & \multicolumn{4}{c||}{ $[1,\textcolor{brown}{2}]_{\textcolor{brown}{10}}$,$[1,\textcolor{brown}{2}]_{\textcolor{brown}{120}}$,$[1,\textcolor{brown}{2}]_{\textcolor{brown}{\overline{126}}}$,$[1,\textcolor{brown}{3}]_{\textcolor{brown}{\overline{126}}}$} &  \\
 higgses & \multicolumn{4}{c||}{ $[1,\textcolor{brown}{2}]_{\textcolor{brown}{126}}$,$[1,\textcolor{brown}{3}]_{\textcolor{brown}{126}}$,$[1,\textcolor{brown}{2}]_{\textcolor{brown}{210'}}$,$[1,\textcolor{brown}{4}]_{\textcolor{brown}{210'}}$,$[1,\textcolor{brown}{2}]_{\textcolor{brown}{320}}$,$[1,\textcolor{brown}{4}]_{\textcolor{brown}{320}}$} &  $\textcolor{brown}{M_{Z}}$\\
    & \multicolumn{4}{c||}{
$[1,\textcolor{brown}{2}]_{\textcolor{brown}{560}}$,$[1,\textcolor{brown}{4}]_{\textcolor{brown}{560}}$,$[1,\textcolor{brown}{2}]_{\textcolor{brown}{\overline{560}}}$,$[1,\textcolor{brown}{4}]_{\textcolor{brown}{\overline{560}}}$} &  \\
\hline\hline Final group      & \multicolumn{4}{c||}{$SU(3)_{c} \oplus U(1)_{EM}$} & $ 0 - \textcolor{brown}{M_{Z}}$\\
\hline\hline
\end{tabular}
\label{tab1}
\end{table*}

From four patterns of $SO(10)$ breaking, that are summarized in the Table \ref{tab1}, we choose two examples, that correspond to two different bases of $X_{\widehat{a}}$ generators (eq. (\ref{case1}) and (\ref{case2})).

For generators $\widehat{Y}$ and $X$ defined by eq. (\ref{case1}) there is only one possible intermediate group - $SU(5) \oplus U(1)_{X}$ (Pattern I in section \ref{SecU1N}). We choose $210$ and $54$ of $SO(10)$ as Higgs multiplets, which leads to the following symmetry breaking chain
\begin{equation}\begin{array}{ll}
\text{Case I:}\hspace{1cm} & SO(10) \stackrel{210}{\longrightarrow} SU(5) \oplus U(1)_{X} \stackrel{24_{54+210}}{\longrightarrow} \\ & \stackrel{24_{54+210}}{\longrightarrow} SU(3)_{c} \oplus SU(2)_{L} \oplus U(1)^{2}\end{array}
\label{chain1}
\end{equation}
Higgs representations in symmetry breaking chains (\ref{chain1}), (\ref{chain2}) and (\ref{chain3}) are denoted in agreement with Table \ref{tab1} and $24_{54+210}$ is a linear combination of $24_{54}$ and $24_{210}$. For $X_{\widehat{a}}$ generators equal to $R$ and $\widehat{B-L}$ (defined by eq. (\ref{case2})) and for two 45-dimensional higgses (the minimal choice) there are two possible intermediate groups - $SU(3)_{c} \oplus SU(2)_{L} \oplus SU(2)_{R} \oplus U(1)_{B-L}$ and $SU(4) \oplus SU(2)_{L} \oplus U(1)_{R}$. The first one (Pattern II in section \ref{SecU1N}) leads to the following symmetry breaking chain
\begin{equation}\begin{array}{l}
\text{Case II:}\\ SO(10) \stackrel{45}{\longrightarrow} SU(3)_{c} \oplus SU(2)_{L} \oplus SU(2)_{R} \oplus U(1)_{B-L} \stackrel{[1,1,3]_{45}}{\longrightarrow} \\  \stackrel{[1,1,3]_{45}}{\longrightarrow} SU(3)_{c} \oplus SU(2)_{L} \oplus U(1)^{2}\end{array}
\label{chain2}
\end{equation}
As mentioned in section \ref{SecInt}, we do not construct full $SO(10)$ models with superpotentials that could realize symmetry breaking chains in Cases I and II. Higgs representations, their masses and breaking chains themselves are all what we need to consider the running of gauge coupling constants with threshold corrections.

A higgs field that breaks the $U(1)^{2}$ group - $\chi$ is also necessary to generate the Majorana mass term for right handed neutrinos $\nu_{R}$. Such a term is needed for the type I see-saw mechanism. The minimal $SO(10)$ representations that could be used for breaking $SU(3)_{c} \oplus SU(2)_{L} \oplus U(1)^{2}$ down to $SU(3)_{c} \oplus SU(2)_{L} \oplus U(1)_{Y}$ are $16$ and $\overline{16}$. However, these representations can generate the Majorana mass term for $\nu_{R}$ only through a non-renormalizable coupling. It's due to $B-L$-charge of $[1,1]_{16}$ and $[1,1]_{\overline{16}}$, which is $1$ and $-1$ respectively. This is the property of almost all SM singlets included in Table \ref{tab1}. Fortunately, fields $[1,1]_{126}$ and $[1,1]_{\overline{126}}$ are exceptional, having $B-L$-charges equal to $2$ and $-2$, respectively. This property allows to introduce renormalizable coupling between $\nu_{R}$ and $\chi$

\begin{equation}
\frac{1}{2}Y_{R}^{*}\,\chi^{*}\,\nu_{R}\nu_{R}\;\;\text{for}\;\;\chi = [1,1]_{\overline{126}}
\end{equation}
This coupling (actually its hermitian conjugation) can be embedded into the $\frac{1}{2}\,Y_{\overline{126}}\,16\,\overline{126}\,16$ unified Yukawa coupling. For this reason, we embed $\chi$ in $\overline{126}$. However, the $126$ of $SO(10)$ is also needed in both Cases. In Case II the $126$ is necessary to include the $\overline{126}$ in the part of superpotential that may generate masses of fields embedded in $\overline{126}$ and the VEV of $\chi$ field. In Case I this part of superpotential may contain $\overline{126}$ without $126$, but the direct mass term $M_{126}(126\,\overline{126})$ is needed to obtain relatively small (a few TeV) mass and VEV of $\chi$ field. Explicit forms of superpotential in both Cases are shown in Appendix \ref{ApW}. The original $\chi$ field, which is embedded in $\overline{126}$, will be denoted by $\chi_{-}$. Analogous field from $126$ will be denoted by $\chi_{+}$. Both $\chi_{+}$ and $\chi_{-}$ could acquire VEVs, which break $G_{3211}$, but the Majorana mass term for $\nu_{R}$ is generated only by $\chi_{-}$. A small Majorana mass term for left-handed neutrinos, that breaks the SM gauge group, may also be generated by VEVs of other fields embedded in $\overline{126}$ and $126$ (type II see-saw).

Standard MSSM Higgs fields ($H_{u}$ and $H_{d}$) can be embedded into $10$ of $SO(10)$, which leads to the $\frac{1}{2}\,Y_{10}\,16\,10\,16$ unified Yukawa coupling. Then, the unified form of all considered $16$-Yukawa couplings is the following

\begin{equation}\frac{1}{2}\,16_{i}\,(Y_{10}^{ij}10+Y_{\overline{126}}^{ij}\overline{126})\,16_{j}\end{equation}
However, $H_{u}$ and $H_{d}$ can be also embedded into $\overline{126}$ and $126$ of $SO(10)$ and in general case, they come from the mixing between $SU(2)_{L}$ doublets contained in $10$, $\overline{126}$ and $126$ of $SO(10)$. It means, that standard Yukawa coupling constants and masses of SM fermions originate from linear combinations of $Y_{10}$ and $Y_{\overline{126}}$ \cite{Joshipura}.

The chosen breaking scenario of $Z'$ gauge symmetry (the same for Cases I and II) is the following
\begin{equation}\begin{array}{l}
SU(3)_{c} \oplus SU(2)_{L} \oplus U(1)^{2} \stackrel{[1,1]_{\overline{126}+126}}{\longrightarrow} \\ \stackrel{[1,1]_{\overline{126}+126}}{\longrightarrow} SU(3)_{c} \oplus SU(2)_{L} \oplus U(1)_{Y} \stackrel{[1,2]_{10+\overline{126}+126}}{\longrightarrow} \\ \stackrel{[1,2]_{10+\overline{126}+126}}{\longrightarrow} SU(3)_{c} \oplus U(1)_{EM}\end{array}
\label{chain3}
\end{equation}
We assume that the $Z'$ boson is accessible to the LHC, having a mass equal to a few GeV. Then $\nu_{R}$ has a similar mass (generated by the same VEV) and type I see-saw is not sufficient to explain small masses of light neutrinos. A possible solution to this problem is small neutrino-Dirac mass term that can be obtained through fine-tuning between two contributions to this mass term, related to $Y_{10}$ and $Y_{\overline{126}}$. One can also try to obtain a fine-tuning in type I+II see-saw \cite{Joshipura} between two contributions to light neutrino masses from both see-saw types. 

Summarizing for the Case I we have chiral superfields in $3\cdot 16$, $210$, $54$, $126$, $\overline{126}$ and $10$ of $SO(10)$. The final field content of Case I is shown in Tables \ref{tab2} and \ref{tab3} in Appendix \ref{ApTab}. The $126$ has been omitted, because it's analogous to $\overline{126}$.

For the Case II we have the chiral $SO(10)$ multiplets: $3\cdot 16$, $2\cdot 45$, $126$, $\overline{126}$ and $10$. Between scales $\mu_{0}$ and $\mu_{1}$ the gauge group is equal to $SU(3)_{c} \oplus SU(2)_{L} \oplus  SU(2)_{R} \oplus U(1)_{\widehat{B-L}}$ with gauge coupling constants $g_{3}$, $g_{2}$, $g_{2R}$ and $g_{\widehat{B-L}}$ respectively. At the scale $\mu_{0}$ these gauge coupling constants are unified

\begin{equation}g_{3}(\mu_{0}) = g_{2}(\mu_{0}) = g_{2R}(\mu_{0}) = g_{\widehat{B-L}}(\mu_{0}) = g_{10}(\mu_{0})\label{unif2h}\end{equation}
At the scale $\mu_{1}$

\begin{equation}g_{R}(\mu_{1}) = g_{2R}(\mu_{1})\label{unif2l}\end{equation}
where $g_{R}$ is the gauge coupling constant of the unbroken $U(1)_{R}$ subgroup of $SU(2)_{R}$. Kinetic terms and the $G_{a b}$ matrix are diagonal in the $(R,\widehat{B-L})$ basis at the $\mu_{1}$ scale

\begin{equation}
G_{a b}= G_{\widehat{a} \widehat{b}} = \left[\begin{array}{cc}
g_{R} & 0 \\
0 & g_{\widehat{B-L}}
\end{array}\right]
\label{Gin12}
\end{equation}
Below this scale the RGE-running dediagonalizes matrices $G_{\widehat{a} \widehat{b}}$ and $\vartheta_{\widehat{a} \widehat{b}}$ as in Case I.

In both Cases these matrices contain additional (with respect to SM) abelian gauge coupling constants that we would like to constrain - $g'_{B-L}$ and $g_{B-L}$, which are defined by the following formula

\begin{equation}\begin{array}{l}
L^{gauge}_{2\,int} \supset \sum_{f} \bar{\Psi}_{f}\gamma_{\mu}\left( X^{T\,f}_{\underline{a}} G_{\underline{a} \underline{b}} A^{\mu}_{\underline{b}}\right)\Psi_{f} = \\ \sum_{f} \bar{\Psi}_{f}\gamma_{\mu}\left(\left[Y^{f},(B-L)^{f}\right]\left[\begin{array}{cc}
 g'  & g'_{B-L} \\
 0 &   g_{B-L}
\end{array}\right]  \left[\begin{array}{c}
B^{0\,\mu} \\
Z'^{0\,\mu}
\end{array}\right]\right)\Psi_{f} \end{array}
\label{L2low}
\end{equation}
As one can see, the basis of abelian gauge bosons $A^{\mu}_{\underline{b}}$ is chosen in such a way that the gauge boson $B^{0}$ couples only to weak hypercharge $Y$ through the $g'$ coupling constant, exactly like in SM. One should remember that $Z'^{0}$ is not the final, physical $Z'$ boson. The latter is a mass eigenstate which is a linear combination of $W^{3}$, $B^{0}$ and $Z'^{0}$. The $\mu_{2}$ scale should be identified with the mass of the $Z'$ boson - $M_{Z'}$. Below this scale the $U(1)^{2}$ symmetry is broken down to $U(1)_{Y}$, so one should integrate out the $Z'$ boson and use the effective theory without $g'_{B-L}$ and $g_{B-L}$.

Using eq. (\ref{L2low}), one can calculate $\vartheta^{-1}_{\underline{a} \underline{b}}$ from its definition which has been done by Zwirner et. al. \cite{Zwirner1}

\begin{equation}
\vartheta^{-1}_{\underline{a} \underline{b}} = 4\pi \left[\begin{array}{cc}
\frac {1}{g'^{2}} & -\frac{g'_{B-L}}{g_{B-L}}\frac {1}{g'^{2}} \\
-\frac{g'_{B-L}}{g_{B-L}}\frac {1}{g'^{2}} & \frac{1}{g_{B-L}^{2}} + \left(\frac{g'_{B-L}}{g_{B-L}}\right)^{2}\frac {1}{g'^{2}}
\end{array}\right]
\label{vart}
\end{equation}
The lowest energy scale for $g'_{B-L}$ and $g_{B-L}$ coupling constants is $\mu_{2} = M_{Z'}$, so we'd like to constrain their values at this particular scale ($g'_{B-L}(M_{Z'})$ and $g_{B-L}(M_{Z'})$). Below this scale the only surviver from the $\vartheta^{-1}_{\underline{a} \underline{b}}$ matrix is $\vartheta^{-1}_{\underline{1} \underline{1}} = \alpha'^{-1} = \frac {1}{g'^{2}}$. This coupling constant then runs down to the $M_{Z}$ scale as in the Standard Model.

The final field content of Case II is shown in Tables \ref{tab5}, \ref{tab8} and \ref{tab6} in Appendix \ref{ApTab}.

\section{Analytic 1-loop RG evolution of gauge coupling constants}{\label{Sec1loop}}

The purpose of this and next sections is to find analytically the region of the two low energy coupling constants ($g'_{B-L}$ and $g_{B-L}$) that is consistent with embedding $Z'$ model into $SO(10)$ and to compare that region with new experimental limits from the LHC. We take into account
\begin{enumerate}
\item two different patterns of $SO(10)$ breaking
\item the possibility of large splittings between the scales $\mu_{0}$, $\mu_{1}$ and $M_{Z'}$
\item potential complexity of the mass spectrum of the representations discussed in the previous sections, however, consistent with various constraints (to be discussed).  
\end{enumerate}
At 1-loop, there are analytic solutions of gauge RGE, so there is no need to choose between the bottom-up and the top-down approach. General solutions of gauge RGE, their boundary conditions and various constraints are collected together and treated as one large system of equations and inequalities. At 2-loop, gauge (and Yukawa) RGE have to be solved numerically, so it's not possible to use our 1-loop approach.

\subsection{Simple decoupling pattern}{\label{SubSecNotresh}}

We begin with 1-loop RG evolutions of the gauge coupling constants under the assumption of a simple decoupling procedure: each non-SM particle is assumed to have a mass exactly equal to one of the scales $\mu_{0}$, $\mu_{1}$, $M_{Z'}$. For the considered spectrum several different arrangements for the mass values are natural. All superpartners in the MSSM, $\nu_{R}$, $\chi_{+}$ and $\chi_{-}$ fields (with their superpartners) are assumed to be "light" and to have masses equal to $M_{Z'}$. All other fields are assumed to be "heavy" with masses equal to the $\mu_{1}$ scale or to the $\mu_{0}$ scale. The reason for this classification comes from the superpotential which is shown in detail in Appendix \ref{ApW} for both cases. It generates effective mass terms with mass parameters proportional to $\mu_{0}$ for most of considered fields. The only exceptions are $15_{54}$ and $\overline{{15}}_{54}$ in Case I (however they have effective mass terms with mass parameters proportional to $\mu_{1}$) and all fields embedded in $16$ of $SO(10)$ (they don't have effective mass terms with mass parameters proportional to $\mu_{1}$). "heavy" fields may have masses that are significantly smaller than $\mu_{0}$ (and closer to $\mu_{1}$) due to the freedom in the values of Yukawa coupling constants that multiply the $\mu_{0}$ VEV in their mass parameters. $15_{54}$ and $\overline{{15}}_{54}$ in Case I may have masses significantly larger than $\mu_{1}$ (and closer to $\mu_{0}$) due to large value of their direct mass parameter - $M_{54}$ in the $M_{54}(54\,54)$ coupling.

From the superpotential, one sees that some "light" fields also have large mass parameters proportional to $\mu_{0}$. These are $H_{u}$, $H_{d}$, $\chi_{+}$, $\chi_{-}$ and their superpartners. They should be "light" because they generate $U(1)^{2}$ breaking ($\chi_{+}$, $\chi_{-}$) or standard electroweak breaking ($H_{u}$, $H_{d}$).\newline Therefore their large mass parameters proportional to $\mu_{0}$ should be canceled with large accuracy (fine-tuning) by other mass parameters (mainly direct ones). For $H_{u}$, $H_{d}$ this is the standard doublet-triplet splitting problem. We have similar additional problem with $\chi_{+}$ and $\chi_{-}$.\newline
Explicit form of superpotentials and mass terms for Case I and Case II are shown in Appendix \ref{ApW}.

We demand two important conditions to be satisfied
\begin{enumerate}
\item $\mu_{0} \leq M_{Pl}$
\item The theory to be perturbative up to $\mu_{0}$ ($g_{10}(\mu_{0}) \leq 4\pi$).
\end{enumerate}
1-loop RGE equations (written explicitly in Appendix \ref{ApRGE}) are solved analytically in three different energy-scale intervals related to three different gauge groups:
\begin{enumerate}
\item Interval $I_{\mu_{1}}^{\mu_{0}}$: $[\mu_{1},\mu_{0}]$ for the intermediate gauge group (Case-dependent)
\item Interval $I_{M_{Z'}}^{\mu_{1}}$: $[M_{Z'},\mu_{1}]$ for the $Z'$ gauge group - \newline $SU(3)_{c} \oplus SU(2)_{L} \oplus U(1)^{2}$
\item Interval $I_{M_{Z}}^{M_{Z'}}$: $[M_{Z},M_{Z'}]$ for the SM gauge group - $SU(3)_{c} \oplus SU(2)_{L} \oplus U(1)_{Y}$
\end{enumerate}
The general form of these solutions for non-abelian (or single $U(1)$) gauge coupling constants is the following:
\begin{equation}
\alpha^{-1}_{A}(\mu_{x}) = \alpha^{-1}_{A}(\mu_{y}) - \frac{b_{A}}{2\pi}\ln\left(\frac{\mu_{x}}{\mu_{y}}\right)
\label{sol1naT0}
\end{equation}
Index $A$ denotes a simple gauge group. $b$ parameters are defined in Appendix \ref{ApRGE}. In general $\mu_{x}$ and $\mu_{y}$ are any two energy scales, at which $\alpha_{A}$ is determined and ($\mu_{x} > \mu_{y}$). In this section they are taken to be equal to ends of one of three considered energy-scale intervals defined above. Analogously, solutions to abelian (more than one $U(1)$) gauge RGE are the following
\begin{equation}
\vartheta^{-1}_{ab}(\mu_{x}) = \vartheta^{-1}_{ab}(\mu_{y}) - \frac{b_{ab}}{2\pi} \ln\left(\frac{\mu_{x}}{\mu_{y}}\right)
\label{sol1aT0}
\end{equation}
Boundary conditions are such that values of gauge coupling constants at ends of energy-scale intervals are appropriately glued with each other. In interval $I_{M_{Z}}^{M_{Z'}}$ gauge coupling constants are converted from $\overline{DR}$ to $\overline{MS}$ regularization scheme. According to \cite{Pokorski}, we include coefficients related to this conversion. Values of $g_{2}$, $g_{3}$ and $g'$ at $M_{Z}$ together with their $1\sigma$ uncertainties are taken from \cite{PDG}. In general we consider three values of $M_{Z'}$ - $2$ TeV, $2.25$ TeV and $2.5$ TeV, but in this section we focus on $M_{Z'} = 2$ TeV. Results for two other values of $M_{Z'}$ are very similar and they were obtained to be compared to experimental constraints that strongly depend on $M_{Z'}$. The final results are regions of two additional gauge coupling constants ($g'_{B-L}(M_{Z'})$ and $g_{B-L}(M_{Z'})$), which are allowed by unification. We follow the convention in \cite{Zwirner1} and normalize these two values to $g_{Z}(M_{Z})=\sqrt{g'^{2}(M_{Z})+g_{2}^{2}(M_{Z})}$.
\begin{equation}\begin{array}{l}
\widetilde{g}'_{B-L}(M_{Z'})= \frac{g'_{B-L}(M_{Z'})}{g_{Z}(M_{Z})} \\ \widetilde{g}_{B-L}(M_{Z'})= \frac{g_{B-L}(M_{Z'})}{g_{Z}(M_{Z})}
\end{array}
\label{Zwircon}
\end{equation}

\subsubsection{Case I}{\label{SubSubSecNotreshI}}

In Case I we consider three subcases related to different choices of masses of "heavy" fields
\begin{enumerate}
\item Case Ia: Only $210$ of $SO(10)$ has mass equal to $\mu_{0}$ and the rest of "heavy" fields have masses equal to $\mu_{1}$.
\item Case Ib: "Heavy" fields from $210$, $\overline{126}$ and $126$ of\newline $SO(10)$ have masses equal to $\mu_{0}$. $54$ of $SO(10)$ and two triplets embedded in $10$ of $SO(10)$ have masses equal to $\mu_{1}$.
\item Case Ic: "Heavy" fields from $210$, $\overline{126}$ and $126$ of $SO(10)$, $15_{54}$ and $\overline{{15}}_{54}$ have masses equal to $\mu_{0}$. Two triplets embedded in $10$ of $SO(10)$ and $24_{54}$ have masses equal to $\mu_{1}$.\end{enumerate}
Values of $b$ parameters in these subcases are different and they are shown in Table \ref{tab0I}.
\begin{table*}[!htbp]
\caption{Values of $b$ parameters in Cases Ia, Ib and Ic.}
\begin{math}\begin{array}{||c||c|c|c|c||c||}
\hline \hline \text{Groups}  & b \text{ parameters}  & \text{Ia} & \text{Ib} & \text{Ic}  & \text{Scales}  \\
\hline\hline \textcolor{violet}{SO(10)} & \left(b_{10}\right) & \multicolumn{3}{c||}{\left(\frac{121}{2}\right)}   &  \textcolor{violet}{\mu_{0}} - M_{Pl} \\
\hline  \textcolor{blue}{SU(5)}\oplus U(1)_{X}  & \left(b_{5}\text{, }b_{X}\right)   &\left(74\text{, }89\right) & \left(4\text{, }24\right) & \left(-3\text{, }12\right) &  \textcolor{blue}{\mu_{1}} - \textcolor{violet}{\mu_{0}}   \\
\hline  SU(3)_{c} \oplus  SU(2)_{L} &  \left(b_{3}\text{, }b_{2}\right)  \left(b_{\widehat{1} \widehat{1}}\text{, }b_{\widehat{1} \widehat{2}}\text{, }b_{\widehat{2} \widehat{2}}\right)  &  \multicolumn{3}{c||}{\left(-3\text{, }1\right) \left(\frac{33}{5}\text{, }-\frac{\sqrt{6}}{5}\text{, }\frac{57}{5}\right) }   &  \textcolor{red}{M_{Z'}} - \textcolor{blue}{\mu_{1}} \\
 \oplus \textcolor{red}{U(1)^{2}} &  \left(b_{\underline{1} \underline{1}}\text{, }b_{\underline{1} \underline{2}}\text{, }b_{\underline{2} \underline{2}}\right) & \multicolumn{3}{c||}{ \left(11\text{, }8\text{, }24\right) } & \\
\hline SU(3)_{c} \oplus  \textcolor{brown}{SU(2)_{L}} & \left(b_{3}\text{, }b_{2}\text{, }b_{Y}\right) & \multicolumn{3}{c||}{\left(-7\text{, }-\frac{19}{6}\text{, }\frac{41}{6}\right) } &  \textcolor{brown}{M_{Z}} - \textcolor{red}{M_{Z'}} \\
\oplus \textcolor{brown}{U(1)_{Y}} & & \multicolumn{3}{c||}{} & \\
\hline\hline
\end{array}\end{math}
\label{tab0I}
\end{table*}
In this section we consider some deviations from the exact unification. Therefore, we assume that the $\mu_{1}$ scale is formally defined by condition - $\alpha_{2}(\mu_{1}) = \vartheta_{\widehat{1} \widehat{1}}(\mu_{1})$. This is the standard choice since $\vartheta_{\widehat{1} \widehat{1}}(\mu_{1})$ is equivalent to $\alpha_{1}(\mu_{1})$ in MSSM.  Eventual deviations from the exact unification are given in eq. (\ref{dev00})
\begin{equation}
\begin{array}{l} D_{1}=\frac{(\vartheta^{-1})_{\widehat{1} \widehat{2}}(\mu_{1})}{(\vartheta^{-1})_{\widehat{1} \widehat{1}}(\mu_{1})}= -\frac{\vartheta_{\widehat{1} \widehat{2}}(\mu_{1})}{\vartheta_{\widehat{2} \widehat{2}}(\mu_{1})}\hspace{0.5cm}
D_{2}=\frac{\alpha_{3}(\mu_{1})-\alpha_{2}(\mu_{1})}{\alpha_{3}(\mu_{1})+\alpha_{2}(\mu_{1})} \\ D_{3}=\frac{\alpha_{5}(\mu_{0})-\alpha_{X}(\mu_{0})}{\alpha_{5}(\mu_{0})+\alpha_{X}(\mu_{0})} \end{array}
\label{dev00}
\end{equation}
$D_{1}$ has different form than $D_{2}$ and $D_{3}$, because it's a deviation from vanishing of only one coupling constant - $\vartheta_{\widehat{1} \widehat{2}}$ and not a deviation from unification of at least two coupling constants. It's also a deviation from diagonal form of the $\vartheta_{\widehat{a} \widehat{b}}(\mu_{1})$ matrix (and its inversion). Therefore, the most natural value that $\vartheta_{\widehat{1} \widehat{2}}(\mu_{1})$ should be small relative to is $\vartheta_{\widehat{1} \widehat{1}}(\mu_{1})$ or $\vartheta_{\widehat{2} \widehat{2}}(\mu_{1})$. The latter possibility is chosen. The formal definition of $\mu_{0}$ scale depends on the kind of running of $\alpha_{5}$ and $ \alpha_{X}$ coupling constants. In the ideal situation $\mu_{0}$ is the scale at which $\alpha_{5} = \alpha_{X}$ and $D_{3} = 0$. However, $\alpha_{5}$ and $\alpha_{X}$ may reach $M_{Pl}$ or the perturbativity limit (at least one of them) before they manage to unify with each other. Then, $\mu_{0}$ is equal to $M_{Pl}$ or to the scale of perturbativity breakdown respectively and  $D_{3} \neq 0$.

One can try to obtain exact unification ($D_{1}=D_{2}=D_{3}=0$). $\widetilde{g}'_{B-L}(M_{Z'})$ and $\widetilde{g}_{B-L}(M_{Z'})$ can be fitted to obtain $D_{1}=D_{3}=0$. One can also obtain $D_{2}=0$ by fitting the value of $M_{Z'}$ (assumed to be equal to masses of all "light" fields). However, this fit gives $M_{Z'} = 61.86$ GeV (for central experimental values of $g_{2}(M_{Z})$, $g_{3}(M_{Z})$ and $g'(M_{Z})$), which is definitely too small. Even the $3\sigma$ deviation from central experimental values of $g_{2}(M_{Z})$, $g_{3}(M_{Z})$ and $g'(M_{Z})$ gives at most $M_{Z'} = 208.42$ GeV, which is still too small to obtain unification in agreement with the recent LHC data. There are no other free parameters that could be fitted this way. $\widetilde{g}'_{B-L}(M_{Z'})$ and $\widetilde{g}_{B-L}(M_{Z'})$ are also free parameters but they have no influence on $D_{2}$. Therefore in simple decoupling pattern in Case I the exact unification is impossible.

One can obtain unification, which is as close to the exact one as possible (deviations as small as possible). Then, $D_{1}=D_{3}=0$, values of $D_{2}$ are equal to $0.01440$, $0.01485$, $0.01525$ and values of $\mu_{1}$ are equal to $8.544\cdot 10^{15}$ GeV, $8.250\cdot 10^{15}$ GeV, $7.995\cdot 10^{15}$ GeV for $M_{Z'}$ equal to $2$ TeV, $2.25$ TeV and $2.5$ TeV respectively (for central experimental values of $g_{2}(M_{Z})$, $g_{3}(M_{Z})$ and $g'(M_{Z})$). These values are exactly the same as in MSSM (without additional $U(1)$ gauge group).

\begin{figure*}
\resizebox{1\textwidth}{!}{
    \includegraphics{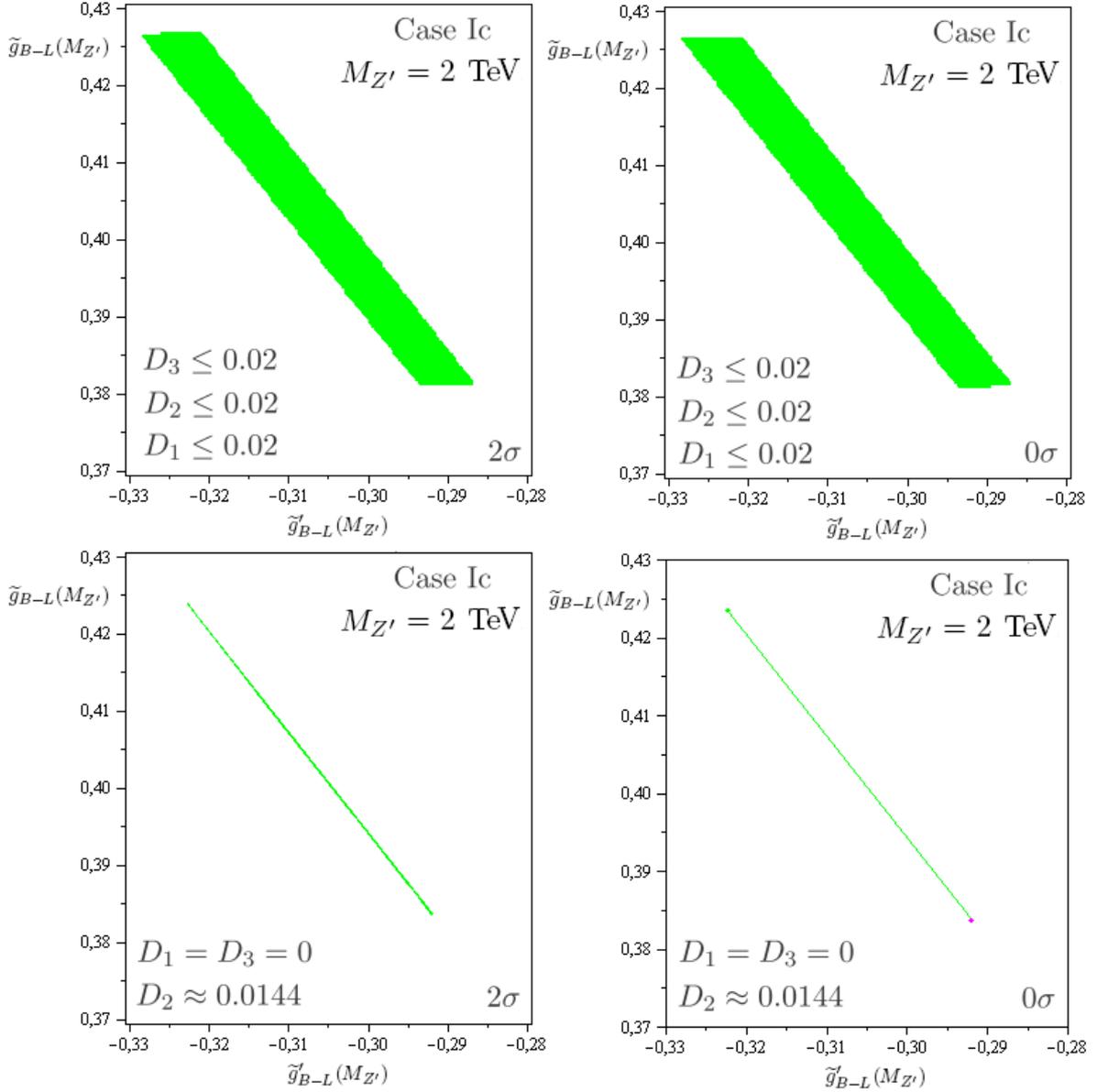}}
  \caption{Values of $g'_{B-L}(M_{Z'})$ and $g_{B-L}(M_{Z'})$ allowed by approximate unification in Case Ic for $M_{Z'} = 2$ TeV. On two upper plots deviations from exact unification ($D_{1}$, $D_{2}$ and $D_{3}$) are smaller than $0.02$, while on two lower plots they are as small as possible. On two left plots values of $g_{2}(M_{Z})$, $g_{3}(M_{Z})$ and $g'(M_{Z})$ differ from central experimental ones no more than $2\sigma$, while on two right plots they are exactly equal to central experimental values. The region shown on the lower right plot is strictly one-dimensional interval.}
  \label{figIc}
\end{figure*}

\begin{figure*}
\resizebox{1\textwidth}{!}{
    \includegraphics{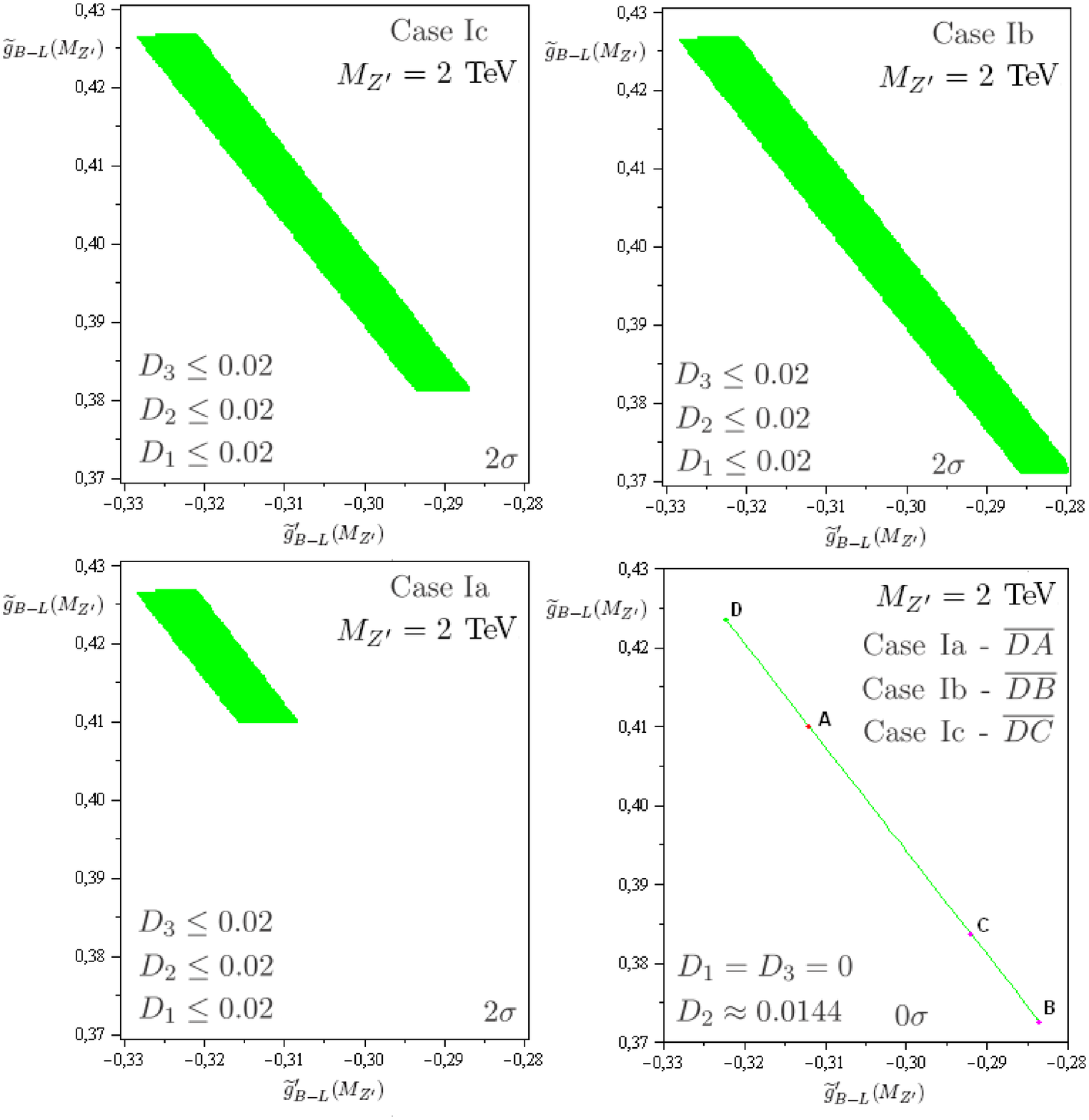}}
  \caption{Values of $g'_{B-L}(M_{Z'})$ and $g_{B-L}(M_{Z'})$ allowed by approximate unification in Cases Ia, Ib and Ic for $M_{Z'} = 2$ TeV. On lower right plot deviations from exact unification ($D_{1}$, $D_{2}$ and $D_{3}$) are as small as possible and values of $g_{2}(M_{Z})$, $g_{3}(M_{Z})$ and $g'(M_{Z})$ are exactly equal to central experimental ones. Regions shown on this plot are strictly one-dimensional intervals. On other plots deviations from exact unification are smaller than $0.02$ and values of $g_{2}(M_{Z})$, $g_{3}(M_{Z})$ and $g'(M_{Z})$ differ from central experimental ones no more than $2\sigma$.}
  \label{figI}
\end{figure*}

The region on the plane spanned by $g'_{B-L}(M_{Z'})$ and $g_{B-L}(M_{Z'})$ that is allowed by approximate unification is shown on Figure \ref{figIc} in Case Ic for $M_{Z'} = 2$ TeV in four cases.
\begin{enumerate}
\item Deviations $D_{1}$, $D_{2}$, $D_{3}$ as small as possible for central experimental values of $g_{2}(M_{Z})$, $g_{3}(M_{Z})$ and $g'(M_{Z})$
\item Deviations $D_{1}$, $D_{2}$, $D_{3}$ as small as possible for values of $g_{2}(M_{Z})$, $g_{3}(M_{Z})$ and $g'(M_{Z})$ that differ from central experimental ones no more than $2\sigma$.
\item Deviations $D_{1}$, $D_{2}$, $D_{3}$ smaller than $0.02$ for central experimental values of $g_{2}(M_{Z})$, $g_{3}(M_{Z})$ and $g'(M_{Z})$
\item Deviations $D_{1}$, $D_{2}$, $D_{3}$ smaller than $0.02$ for values of $g_{2}(M_{Z})$, $g_{3}(M_{Z})$ and $g'(M_{Z})$ that differ from central experimental ones no more than $2\sigma$.
\end{enumerate}
One can see that $2\sigma$ deviations from central experimental values of $g_{2}(M_{Z})$, $g_{3}(M_{Z})$ and $g'(M_{Z})$ are negligible when compared to allowed deviations from exact unification. It's also true for Cases Ia and Ib and for $M_{Z'}$ equal to $2.25$ TeV and $2.5$ TeV.
On Figure \ref{figI} we show results for all three Cases - Ia Ib and Ic.
When deviations $D_{1}$, $D_{2}$, $D_{3}$ are as small as possible, for central experimental values of $g_{2}(M_{Z})$, $g_{3}(M_{Z})$ and $g'(M_{Z})$, one obtains strictly one-dimensional interval allowed for $g'_{B-L}(M_{Z'})$ and $g_{B-L}(M_{Z'})$, because there is one equality that\newline $g'_{B-L}(M_{Z'})$ and $g_{B-L}(M_{Z'})$ have to satisfy and (for $M_{Z'}$ fixed) there are no more independent free parameters. The equality originates from the $\vartheta_{\widehat{1} \widehat{2}}(\mu_{1})=0$ unification condition (see eq.(\ref{Gin1})), which is equivalent to $D_{1}=0$. One end of the allowed interval corresponds to the $\mu_{1}=\mu_{0}$ limit. Three subcases (Ia, Ib and Ic) differ from each other only in the $I_{\mu_{1}}^{\mu_{0}}$ interval. Therefore in the $\mu_{1}=\mu_{0}$ limit they are identical and on Figure \ref{figI} there is only one point for all of them - point D. For the same reason the $\vartheta_{\widehat{1} \widehat{2}}(\mu_{1})=0$ condition is identical for all three subcases, so three intervals on Figure \ref{figI} are parallel to each other. For Cases Ib and Ic the other end of the allowed interval corresponds to the $\mu_{0} = M_{Pl}$ limit and for the Case Ia it corresponds to the perturbativity limit ($g_{10}(\mu_{0}) = 4\pi$). In Cases Ib and Ic the relatively small number of fields with masses equal to $\mu_{1}$ causes that the running of gauge coupling constants above $\mu_{1}$ is slow enough that they can reach the $M_{Pl}$ scale before the perturbativity breakdown. It's not true in Case Ia.

\subsubsection{Case II}{\label{SubSubSecNotreshII}}
 
In Case II we also consider three subcases related to different choices of masses of "heavy" fields
\begin{enumerate}
\item Case IIa: Only one $45$ of $SO(10)$ has mass equal to $\mu_{0}$ and the rest of "heavy" fields have masses equal to $\mu_{1}$.
\item Case IIb: $\mu_{0}$ is equal to the mass of one $45$ of $SO(10)$ and six $SU(3)_{c}$ triplets from $10$, $\overline{126}$ and $126$ that can mediate proton decay. The rest of "heavy" fields - the other $45$ of $SO(10)$ and remaining "heavy" parts of $\overline{126}$ and $126$ have masses equal to $\mu_{1}$.
\item Case IIc: $\mu_{1}$ is equal to the mass of "heavy" fields in two $SU(2)_{R}$ triplets from $\overline{126}$ and $126$, that contain $\chi_{-}$ and $\chi_{+}$ respectively. It's also equal to the mass of another $SU(2)_{R}$ triplet from $45$ of $SO(10)$ that breaks the $SU(2)_{R}$ group. The rest of "heavy" fields have masses equal to $\mu_{0}$.
\end{enumerate}
Values of $b$ parameters in these subcases are different and they are shown in Table\ref{tab0II}.
\begin{table*}[!htbp]
\caption{Values of $b$ parameters in Cases IIa, IIb and IIc.}
\begin{math}\begin{array}{||c||c|c|c|c||c||}
\hline \hline \text{Groups}  & b \text{ parameters}  & \text{IIa} & \text{IIb} & \text{IIc}  & \text{Scales}  \\
\hline\hline \textcolor{violet}{SO(10)} & \left(b_{10}\right) & \multicolumn{3}{c||}{\left(\frac{69}{2}\right)}   &  \textcolor{violet}{\mu_{0}} - M_{Pl} \\
\hline  SU(3)_{c}\oplus SU(2)_{L}  & \left(b_{3}\text{, }b_{2}\right)   &\left(76\text{, }79\right) & \left(73\text{, }79\right) & \left(-3\text{, }1\right) &  \textcolor{blue}{\mu_{1}} - \textcolor{violet}{\mu_{0}}   \\
  \oplus\textcolor{blue}{SU(2)_{R}\oplus U(1)_{B-L}} & \left(b_{2\,R}\text{, }b_{B-L}\right)   &\left(79\text{, }85\right) & \left(79\text{, }82\right) & \left(7\text{, }15\right) &   \\
\hline  SU(3)_{c} \oplus  SU(2)_{L} &  \left(b_{3}\text{, }b_{2}\right)  \left(b_{\widehat{1} \widehat{1}}\text{, }b_{\widehat{1} \widehat{2}}\text{, }b_{\widehat{2} \widehat{2}}\right)  &  \multicolumn{3}{c||}{\left(-3\text{, }1\right) \left(9\text{, }-\sqrt{6}\text{, }9\right) }   &  \textcolor{red}{M_{Z'}} - \textcolor{blue}{\mu_{1}} \\
 \oplus \textcolor{red}{U(1)^{2}} &  \left(b_{\underline{1} \underline{1}}\text{, }b_{\underline{1} \underline{2}}\text{, }b_{\underline{2} \underline{2}}\right) & \multicolumn{3}{c||}{ \left(11\text{, }8\text{, }24\right) } & \\
\hline SU(3)_{c} \oplus  \textcolor{brown}{SU(2)_{L}} & \left(b_{3}\text{, }b_{2}\text{, }b_{Y}\right) & \multicolumn{3}{c||}{\left(-7\text{, }-\frac{19}{6}\text{, }\frac{41}{6}\right) } &  \textcolor{brown}{M_{Z}} - \textcolor{red}{M_{Z'}} \\
\oplus \textcolor{brown}{U(1)_{Y}} & & \multicolumn{3}{c||}{} & \\
\hline\hline
\end{array}\end{math}
\label{tab0II}
\end{table*}
The $\mu_{1}$ scale is defined by the only possible condition - $\vartheta_{\widehat{1} \widehat{2}}(\mu_{1})=0$. Eventual deviations from the exact unification are given in eq. (\ref{dev01})
\begin{equation} 
\begin{array}{l} D'_{1}=\frac{\alpha_{2R}(\mu_{0})-\alpha_{\widehat{B-L}}(\mu_{0})}{\alpha_{2R}(\mu_{0})+\alpha_{\widehat{B-L}}(\mu_{0})}\hspace{0.5cm}
D'_{2}=\frac{\alpha_{3}(\mu_{0})-\alpha_{2}(\mu_{0})}{\alpha_{3}(\mu_{0})+\alpha_{2}(\mu_{0})} \\ D'_{3}=\frac{\alpha_{2}(\mu_{0})-\alpha_{\widehat{B-L}}(\mu_{0})}{\alpha_{2}(\mu_{0})+\alpha_{\widehat{B-L}}(\mu_{0})} \end{array}
\label{dev01}
\end{equation}
The formal definition of $\mu_{0}$ scale depends on the kind of running of four gauge coupling constants above the $\mu{1}$ scale. In the ideal situation $\mu_{0}$ is the scale at which $\alpha_{\widehat{B-L}} = \alpha_{2}$ and $D'_{3} = 0$. However, gauge coupling constants may reach $M_{Pl}$ or the perturbativity limit (at least one of them) before $\alpha_{\widehat{B-L}}$ and $\alpha_{2}$ manage to unify with each other. Then, $\mu_{0}$ is equal to $M_{Pl}$ or to the scale of perturbativity breakdown respectively and $D'_{3} \neq 0$.

\begin{figure*}
\resizebox{1\textwidth}{!}{
    \includegraphics{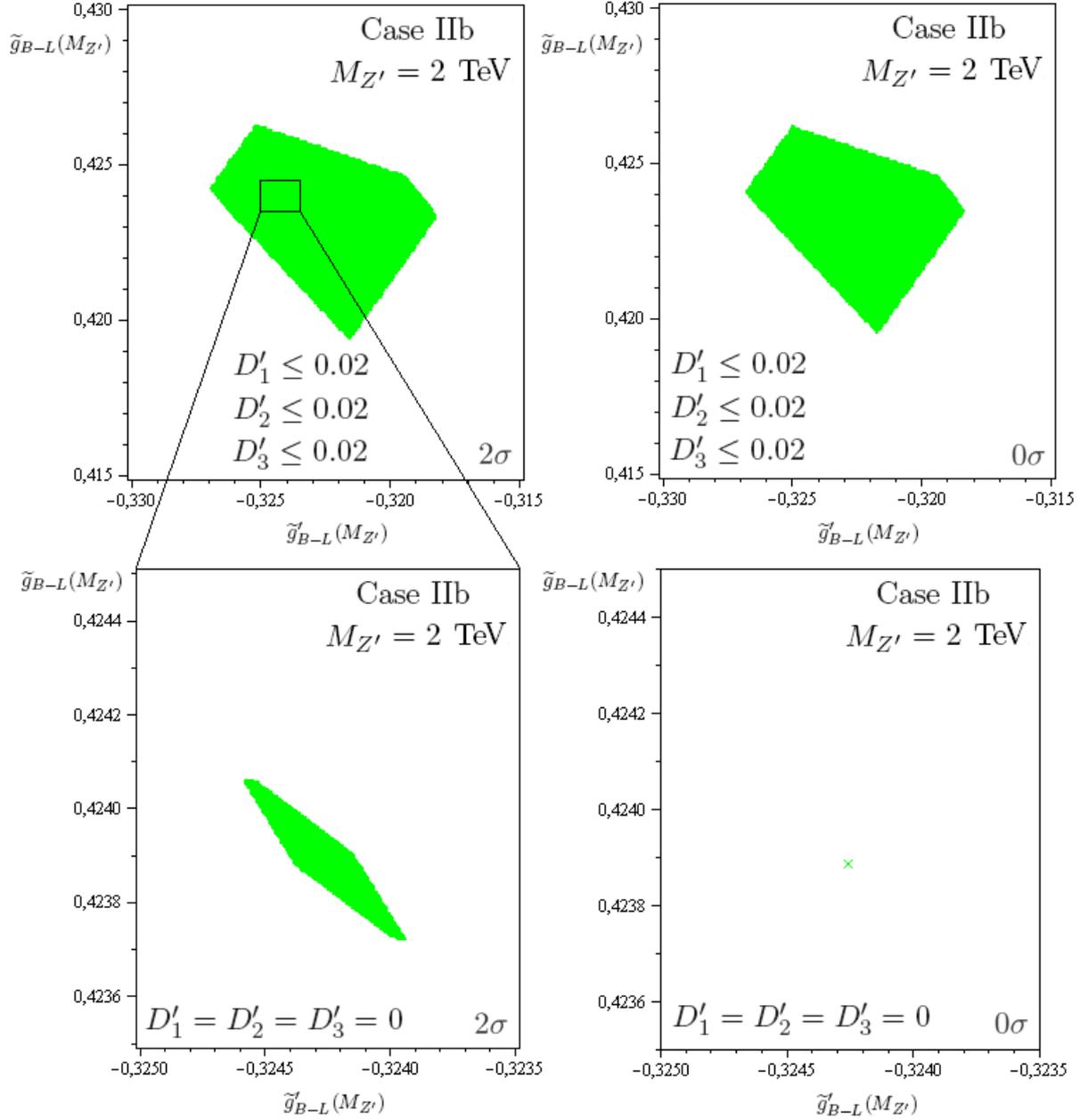}}
  \caption{Values of $g'_{B-L}(M_{Z'})$ and $g_{B-L}(M_{Z'})$ allowed by (approximate) unification in Case IIb for $M_{Z'} = 2$ TeV. On two upper plots deviations from exact unification ($D_{1}$, $D_{2}$ and $D_{3}$) are smaller than $0.02$, while on two lower plots they are equal to $0$ - unification is exact. On two left plots values of $g_{2}(M_{Z})$, $g_{3}(M_{Z})$ and $g'(M_{Z})$ differ from central experimental ones no more than $2\sigma$, while on two right plots they are exactly equal to central experimental values. Only one point is allowed in case shown on the lower right plot. The region shown on the lower left plot corresponds to the small rectangle shown on the upper left plot.}
  \label{figIIb}
\end{figure*}

\begin{figure*}
\resizebox{1\textwidth}{!}{
    \includegraphics{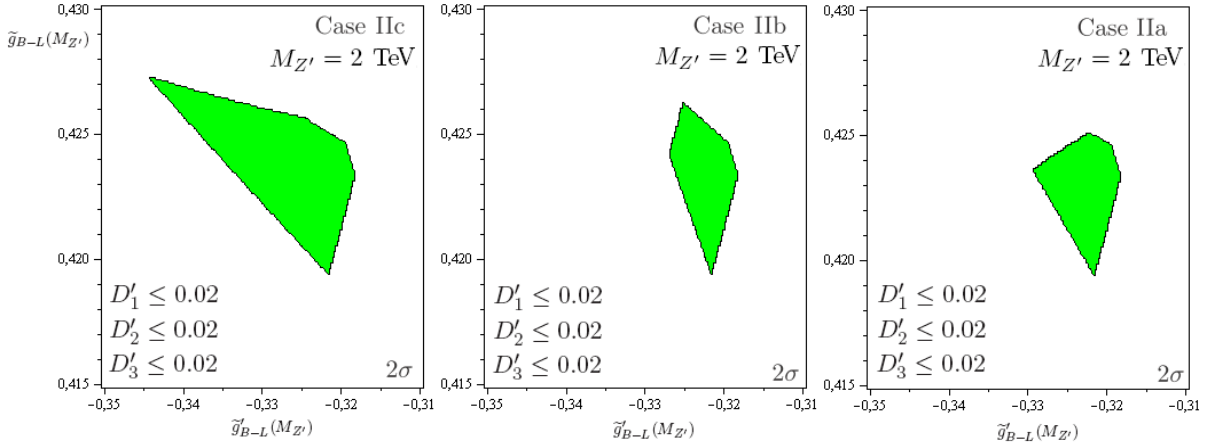}}
  \caption{Values of $g'_{B-L}(M_{Z'})$ and $g_{B-L}(M_{Z'})$ allowed by approximate unification in Cases IIa, IIb and IIc for $M_{Z'} = 2$ TeV. Deviations from exact unification are smaller than $0.02$ and values of $g_{2}(M_{Z})$, $g_{3}(M_{Z})$ and $g'(M_{Z})$ differ from central experimental ones no more than $2\sigma$.}
  \label{figII}
\end{figure*}

We choose the Case IIb for $M_{Z'}=2$ TeV for detailed analysis of the influence of deviations from unification and deviations from central experimental values of $g_{2}(M_{Z})$, $g_{3}(M_{Z})$ and $g'(M_{Z})$ coupling constants on the allowed region of $g'_{B-L}(M_{Z'})$ and $g_{B-L}(M_{Z'})$ coupling constants. For exact unification ($D'_{1}=D'_{2}=D'_{3}=0$), for central experimental values of $g_{2}(M_{Z})$, $g_{3}(M_{Z})$ and $g'(M_{Z})$ this region is reduced to only one point. The 1-loop running related to this point gives $\mu_{1}=7.129\cdot10^{15}$ GeV and $\mu_{0}=1.806\cdot10^{16}$ GeV. It's a point and not one-dimensional interval, because in addition to the $\vartheta_{\widehat{1} \widehat{2}}(\mu_{1})=0$ condition eq. (\ref{unif2h}) must also be satisfied and they give two equalities on $g'_{B-L}(M_{Z'})$ and $g_{B-L}(M_{Z'})$. Obtained results are shown on Figure \ref{figIIb}
As in Case I, $2\sigma$ deviations from central experimental values of $g_{2}(M_{Z})$, $g_{3}(M_{Z})$ and $g'(M_{Z})$ are negligible when compared to allowed deviations from exact unification. It's also true for Cases IIa and IIc and for $M_{Z'}$ equal to $2.25$ TeV and $2.5$ TeV. On Figure \ref{figII} we show results for all three Cases - IIa IIb and IIc.

\subsection{1-loop RGE with mass splitting}{\label{SubSecTresh}}

In this subsection we consider the case in which masses of non-SM particles are not necessarily equal to symmetry breaking scales. When compared to the previous subsection, this gives additional freedom which may help to obtain unification. Therefore, only exact unification is considered here. Moreover, values of $g_{2}(M_{Z})$, $g_{3}(M_{Z})$ and $g'(M_{Z})$ are strictly equal to central experimental ones. Two conditions from the previous subsection -  $\mu_{0} \leq M_{Pl}$ and perturbativity up to $\mu_{0}$ remain unchanged. We introduce one more condition to avoid too rapid proton decay \cite{Senjanovic}:

The unification scale of a GUT gauge group that allows for proton decay has to be larger than $~10^{16}$ GeV. Therefore, in Case I $\mu_{1} > 10^{16}$ GeV and in Case II $\mu_{0} > 10^{16}$ GeV, because in Case II the intermediate gauge group doesn't allow for proton decay. In this Case it's allowed only by $SO(10)$ and $\mu_{1}$ can be much smaller than $10^{16}$ GeV.

Due to the large number of non-SM particles, it's not convenient to include a threshold correction in each gauge RGE for each of these particles explicitly. Instead one can introduce effective threshold corrections related to gauge coupling constants with effective threshold mass parameters $M_{A}$ and $M_{ab}$ defined by equations (\ref{treshna}) and (\ref{tresha}) respectively.

\begin{equation}\begin{array}{l}
(b^{\mu_{x}}_{A}-b^{\mu_{y}}_{A})\ln\left(\frac{M_{A}}{\mu_{y}}\right)=\sum_{f:\mu_{y}<m_{f}<\mu_{x}}b^{f}_{A}\ln\left(\frac{m_{f}}{\mu_{y}}\right) +\\+ \sum_{s:\mu_{y}<m_{s}<\mu_{x}}b^{s}_{A}\ln\left(\frac{m_{s}}{\mu_{y}}\right)\end{array}
\label{treshna}
\end{equation}
Index $A$ denotes a simple gauge group as in eq. (\ref{sol1naT0}). There is no summation over this index in the left-hand side of eq. (\ref{treshna}). There are summations over only these fields that have physical masses $m_{f}$ and $m_{s}$ between energy scales $\mu_{x}$ and $\mu_{y}$.
$b^{f}_{A}$ and $b^{s}_{A}$ are contributions to $b_{A}$ coming from fields with indices $f$ (fermions) and $s$ (scalars) respectively. $b^{\mu_{x}}_{A}$ is the $b_{A}$ parameter at the scale $\mu_{x}$ which means that it includes only particles with physical masses smaller than $\mu_{x}$.
\begin{equation}
b^{\mu_{x}}_{A} = b_{C_{2}(A)} + \sum_{f:m_{f}<\mu_{x}}b^{f}_{A} + \sum_{s:m_{s}<\mu_{x}}b^{s}_{A}
\label{simpbA1-loop}
\end{equation}
$b_{C_{2}(A)}$ is proportional to the quadratic Casimir operator - $C_{2}(A)$ of the group $A$. Formula (\ref{simpbA1-loop}) is an example of the formula (\ref{bA1-loop}) (in Appendix \ref{ApRGE}), in which $b_{C_{2}(A)}$, $b^{f}_{A}$ and $b^{s}_{A}$ parameters are replaced by explicit expressions (definitions of these parameters). In particular:
\begin{equation}
b^{\mu_{x}}_{A}-b^{\mu_{y}}_{A} = \sum_{f:\mu_{y}<m_{f}<\mu_{x}}b^{f}_{A} + \sum_{s:\mu_{y}<m_{s}<\mu_{x}}b^{s}_{A}
\end{equation}
Equation (\ref{tresha}) is the analog of eq. (\ref{treshna}) for abelian gauge coupling constants

\begin{equation}\begin{array}{l}
(b^{\mu_{x}}_{ab}-b^{\mu_{y}}_{ab})\ln\left(\frac{M_{ab}}{\mu_{y}}\right)=\sum_{f:\mu_{y}<m_{f}<\mu_{x}}b^{f}_{ab}\ln\left(\frac{m_{f}}{\mu_{y}}\right) + \\ \sum_{s:\mu_{y}<m_{s}<\mu_{x}}b^{s}_{ab}\ln\left(\frac{m_{s}}{\mu_{y}}\right)\end{array}
\label{tresha}
\end{equation}
There is no summation over indices $a,b$ in the left-hand side of eq. (\ref{tresha}) and all matrices in this equation are symmetric. One can transform eq. (\ref{treshna}) and (\ref{tresha}) to see that effective mass parameters $M_{A}$ and $M_{ab}$ are actually geometrical, weighted averages of physical masses $m_{f}$ and $m_{s}$ with weights being proportional to $b^{f}_{A}$, $b^{s}_{A}$, $b^{f}_{ab}$ and $b^{s}_{ab}$ parameters:

\begin{equation}
M_{A}=\left(\prod_{f:\mu_{y}<m_{f}<\mu_{x}}m_{f}^{b^{f}_{A}}\cdot\prod_{s:\mu_{y}<m_{s}<\mu_{x}}m_{s}^{b^{s}_{A}}\right)^{\frac{1}{b^{\mu_{x}}_{A}-b^{\mu_{y}}_{A}}}
\label{treshnaG}
\end{equation}
\begin{equation}
M_{ab}=\left(\prod_{f:\mu_{y}<m_{f}<\mu_{x}}m_{f}^{b^{f}_{ab}}\cdot\prod_{s:\mu_{y}<m_{s}<\mu_{x}}m_{s}^{b^{s}_{ab}}\right)^{\frac{1}{b^{\mu_{x}}_{ab}-b^{\mu_{y}}_{ab}}}
\label{treshaG}
\end{equation}
With effective threshold corrections, solutions to\newline non-abelian (or single $U(1)$) gauge RGE are the following

\begin{equation}
\alpha^{-1}_{A}(\mu_{x}) = \alpha^{-1}_{A}(\mu_{y}) - \frac{b^{\mu_{x}}_{A}}{2\pi}\ln\left(\frac{\mu_{x}}{\mu_{y}}\right) + \frac{b^{\mu_{x}}_{A}-b^{\mu_{y}}_{A}}{2\pi}\ln\left(\frac{M_{A}}{\mu_{y}}\right)
\label{sol1naT}
\end{equation}
Analogously, solutions to abelian (more than one $U(1)$) gauge RGE are the following

\begin{equation}
\vartheta^{-1}_{ab}(\mu_{x}) = \vartheta^{-1}_{ab}(\mu_{y}) - \frac{b^{\mu_{x}}_{ab}}{2\pi} \ln\left(\frac{\mu_{x}}{\mu_{y}}\right) + \frac{b^{\mu_{x}}_{ab}-b^{\mu_{y}}_{ab}}{2\pi}\ln\left(\frac{M_{ab}}{\mu_{y}}\right)
\label{sol1aT}
\end{equation}
We assumed that masses of all "heavy" fields belong to the $[\mu_{1},\mu_{0}]$ range. Therefore, these masses are replaced by effective threshold mass parameters $M_{A}$ related to simple subgroups of the (Case-dependent) intermediate gauge group related to the Interval $I_{\mu_{1}}^{\mu_{0}}$. Eq. (\ref{treshnaG}) and\newline non-negative values of all $b$ parameters of all "heavy" fields imply that the range of values of these $M_{A}$ threshold parameters is just equal to $[\mu_{1},\mu_{0}]$. These threshold parameters are considered in more detail in next two subsections.

"Light" fields cannot be treated in such a simple way. If we assume that the $Z'$ boson is light enough to be detectable in the LHC, then its mass could be comparable with masses of many "light" (mainly MSSM) particles. In particular, some of them could be heavier than $Z'$ boson and others could be lighter. Since we don't know the spectrum of the MSSM, there are plenty of possible ways to divide these particles into these two categories. For different ways, we have different values of $b^{M_{Z'}}_{A}$ and $b^{M_{Z'}}_{ab}$ - parameters in eq. (\ref{sol1naT}) and (\ref{sol1aT}) in Intervals $I_{M_{Z'}}^{\mu_{1}}$ and $I_{M_{Z}}^{M_{Z'}}$.

Fortunately, there is a solution to this problem that collects all those possibilities to just two ones. First we assume that masses of all "light" fields belong to the $[1,10]$ TeV range. In Case II, where $\mu_{1}$ can be much smaller than $10^{16}$ GeV, we assume that it's larger than $10$ TeV so every "heavy" field is indeed heavier than every "light" field. As in the previous subsection we consider three values of $M_{Z'}$ - $2$ TeV, $2.25$ TeV and $2.5$ TeV - all inside the $[1,10]$ TeV range. Then, we substitute $\mu_{y} = 1$ TeV and $\mu_{x} = 10$ TeV to formulas (\ref{treshnaG}) and (\ref{treshaG}) and use them to calculate the allowed ranges for the following effective threshold mass parameters $M_{2}$, $M_{3}$, $M_{\underline{1} \underline{1}}$, $M_{\underline{1} \underline{2}}$ and $M_{\underline{2} \underline{2}}$ for (inverted) gauge coupling constants $\alpha_{2}^{-1}$, $\alpha_{3}^{-1}$, $\vartheta^{-1}_{\underline{1} \underline{1}} = \alpha'^{-1}$, $\vartheta^{-1}_{\underline{1} \underline{2}}$ and $\vartheta^{-1}_{\underline{2} \underline{2}}$ respectively. The allowed ranges are $[1,10]$ TeV for $M_{2}$, $M_{3}$, $M_{\underline{1} \underline{1}}$, $M_{\underline{2} \underline{2}}$ and $[10^{-\frac{1}{8}}, 10^{\frac{9}{8}}]$ TeV $\approx [0.75, 13.3]$ TeV for $M_{\underline{1} \underline{2}}$. The range for $M_{\underline{1} \underline{2}}$ is exceptional, because $b^{\widetilde{d}}_{\underline{1}\underline{2}} = -\frac{1}{9} < 0$ and so $M_{\underline{1} \underline{2}}$ is a decreasing function of three $\widetilde{d}$ squark masses. All other $b^{f}$ and $b^{s}$ parameters in formulas (\ref{treshnaG}) and (\ref{treshaG}) are non-negative. The next step is to distinguish one of above five effective threshold mass parameters and treat it as the general average mass scale for all "light" fields. For technical reasons, we've chosen $M_{\underline{1} \underline{1}}$. Since it depends only on the (non-SM) MSSM fields (weak hypercharges of other "light" fields are equal to $0$), it will be also denoted by $T_{SUSY}$. From eq. \ref{treshaG} we obtain:
\begin{equation}\begin{array}{l}
T_{SUSY} = M_{\underline{1} \underline{1}}=\left(\prod_{f:1\,\text{TeV}<m_{f}<10\,\text{TeV}}m_{f}^{b^{f}_{\underline{1} \underline{1}}}\cdot\right. \\ \left. \cdot\prod_{s:1\,\text{TeV}<m_{s}<10\,\text{TeV}}m_{s}^{b^{s}_{\underline{1} \underline{1}}}\right)^{\frac{1}{b^{10\,\text{TeV}}_{\underline{1} \underline{1}}-b^{1\,\text{TeV}}_{\underline{1} \underline{1}}}}\end{array}
\label{treshaTSUSY}
\end{equation}
The four remaining threshold mass parameters - $M_{2}$, $M_{3}$, $M_{\underline{1} \underline{2}}$ and $M_{\underline{2} \underline{2}}$ can be replaced by "jumps" of gauge coupling constants at the $T_{SUSY}$ scale. These "jumps" will be denoted by $s_{2}$, $s_{3}$, $s_{\underline{1} \underline{2}}$ and $s_{\underline{2} \underline{2}}$ respectively and they are defined in the following way
\begin{equation}
s_{A} = \frac{b^{10\,\text{TeV}}_{A}-b^{1\,\text{TeV}}_{A}}{2\pi}\ln\left(\frac{M_{A}}{T_{SUSY}}\right)\text{,  where } A\in{2,3} 
\label{jumps}
\end{equation}
\begin{equation}
s_{\underline{a} \underline{b}} = \frac{b^{10\,\text{TeV}}_{\underline{a} \underline{b}}-b^{1\,\text{TeV}}_{\underline{a} \underline{b}}}{2\pi}\ln\left(\frac{M_{\underline{a} \underline{b}}}{T_{SUSY}}\right)
\label{jumps2}
\end{equation}
Allowed ranges for $M_{2}$, $M_{3}$, $M_{\underline{1} \underline{2}}$ and $M_{\underline{2} \underline{2}}$ are transformed to appropriate ranges for "jumps". Under our assumptions $b^{1\,\text{TeV}}_{A} = b^{M_{Z}}_{A}$, $b^{10\,\text{TeV}}_{A} = b^{\mu_{1}}_{A}$, $b^{1\,\text{TeV}}_{\underline{a} \underline{b}} = b^{M_{Z}}_{\underline{a} \underline{b}}$ and $b^{10\,\text{TeV}}_{\underline{a} \underline{b}} = b^{\mu_{1}}_{\underline{a} \underline{b}}$, because there are no non-SM fields with masses in the $[M_{Z},1\,\text{TeV}]$ or in the $[10\,\text{TeV},\mu_{1}]$ range.

Then there are only two major cases to consider.
\begin{enumerate}
\item Case A: $1$ TeV $< M_{Z'} < T_{SUSY} < 10$ TeV with four "jumps" - $s_{2}$, $s_{3}$, $s_{\underline{1} \underline{2}}$ and $s_{\underline{2} \underline{2}}$. In this case we have the following four intervals to consider
\begin{enumerate}
\item Interval $I_{\mu_{1}}^{\mu_{0}}$ - the intermediate gauge group (different in Case I and Case II)
\item Interval $I_{T_{SUSY}}^{\mu_{1}}$: $[T_{SUSY},\mu_{1}]$ - the $G_{3211}$ gauge\newline group and RGE running with $b^{\mu_{1}}_{A}$ and $b^{\mu_{1}}_{\underline{a} \underline{b}}$ parameters.
\item Interval $I_{M_{Z'}}^{T_{SUSY}}$: $[M_{Z'},T_{SUSY}]$ - the $G_{3211}$ gauge group and RGE running with $b^{M_{Z}}_{A}$ and $b^{M_{Z}}_{\underline{a} \underline{b}}$ parameters.
\item Interval $I_{M_{Z}}^{M_{Z'}}$ - the SM gauge group and RGE running with $b^{M_{Z}}_{A}$ parameters ($A=1$ for $U(1)_{Y}$).
\end{enumerate}
In this case values of $b^{M_{Z'}}_{A}$ and $b^{M_{Z'}}_{\underline{a} \underline{b}}$ are equal to $b^{M_{Z}}_{A}$ and $b^{M_{Z}}_{\underline{a} \underline{b}}$.
\item Case B: $1$ TeV $< T_{SUSY} < M_{Z'} < 10$ TeV with only two "jumps" - $s_{2}$ and $s_{3}$. In this case we have the following four intervals to consider
\begin{enumerate}
\item Interval $I_{\mu_{1}}^{\mu_{0}}$ - the intermediate gauge group (different in Case I and Case II)
\item Interval $I_{M_{Z'}}^{\mu_{1}}$ - the $G_{3211}$ gauge group and RGE running with $b^{\mu_{1}}_{A}$ and $b^{\mu_{1}}_{\underline{a} \underline{b}}$ parameters.
\item Interval $I_{T_{SUSY}}^{M_{Z'}}$: $[T_{SUSY},M_{Z'}]$ - the SM gauge\newline group and RGE running with $b^{\mu_{1}}_{A}$ parameters.
\item Interval $I_{M_{Z}}^{T_{SUSY}}$: $[M_{Z},T_{SUSY}]$ - the SM gauge group and RGE running with $b^{M_{Z}}_{A}$ parameters.
\end{enumerate}
In this case values of $b^{M_{Z'}}_{A}$ and $b^{M_{Z'}}_{\underline{a} \underline{b}}$ are equal to $b^{\mu_{1}}_{A}$ and $b^{\mu_{1}}_{\underline{a} \underline{b}}$. Since $\vartheta^{-1}_{\underline{1} \underline{2}}$ and $\vartheta^{-1}_{\underline{2} \underline{2}}$ are running only above $M_{Z'}$ scale, they cannot be corrected by $s_{\underline{1} \underline{2}}$ and $s_{\underline{2} \underline{2}}$ at lower $T_{SUSY}$ scale. It means that in Case B these two gauge coupling constants are not threshold-corrected at all which is sort of approximation in our approach. Even if $T_{SUSY} < M_{Z'}$, some "light" particles may be heavier than $Z'$ and they should give threshold corrections to $\vartheta^{-1}_{\underline{1} \underline{2}}$ and $\vartheta^{-1}_{\underline{2} \underline{2}}$. 
\end{enumerate}

Cases A and B will be considered in Cases I and II, so in total we have four subcases that will be denoted naturally by IA, IB, IIA and IIB. Generally capital letters denote subcases with mass splittings, while small letters denote subcases in simple decoupling pattern (Ia, Ib, Ic, IIa, IIb and IIc in previous subsection). 

Having the field content of the theory, solutions of gauge RGE equations with effective threshold corrections, unification relations for gauge coupling constants and constraining inequalities one can combine all these information, eliminate unwanted variables and finally obtain constraints on coupling constants $g'_{B-L}(M_{Z'})$ and\newline $g_{B-L}(M_{Z'})$.
These constraints for all Cases are presented in the following subsections. The elimination of unwanted variables starts from equations. We use all equations to eliminate as many variables as possible. Then we are left with inequalities that still contain unwanted variables. In the space of variables one can easily find a base in which all inequalities are linear and define the allowed multidimensional polyhedron in this space. We analytically project this polyhedron on the subspace spanned by $g'_{B-L}(M_{Z'})$ and $g_{B-L}(M_{Z'})$ only in Case A and on the subspace spanned by\newline $g'_{B-L}(M_{Z'})$, $g_{B-L}(M_{Z'})$ and $T_{SUSY}$ in Case B. In the latter case $T_{SUSY}$ cannot be eliminated as easily as other unwanted parameters, because remaining inequalities contain non-linear dependence between $T_{SUSY}$ and the\newline $\frac{g'_{B-L}(M_{Z'})}{g_{B-L}(M_{Z'})}$ ratio. However, in Case B the final elimination of $T_{SUSY}$ from complicated, non-linear set of inequalities containing only three variables - $g'_{B-L}(M_{Z'})$, $g_{B-L}(M_{Z'})$ and $T_{SUSY}$ can still be done analytically.

\subsubsection{Case I: Breaking \texorpdfstring{$SO(10) \rightarrow SU(5) \oplus U(1)_{X}$}{\space}}{\label{SubSubSecTreshI}}

There are two gauge coupling constants related to the intermediate group $\rightarrow SU(5) \oplus U(1)_{X}$ - $g_{5}$ and $g_{X}$ respectively. $m_{5}$ and $m_{X}$ are related mass parameters of effective threshold corrections caused by "heavy" fields. Using eq. \ref{treshnaG} we obtain:
\begin{equation}
m_{5}=\left(\prod_{f:\mu_{1}<m_{f}<\mu_{0}}m_{f}^{b^{f}_{5}}\cdot\prod_{s:\mu_{1}<m_{s}<\mu_{0}}m_{s}^{b^{s}_{5}}\right)^{\frac{1}{b^{\mu_{1}}_{5}-b^{\mu_{0}}_{5}}}
\label{treshnaGm5}
\end{equation}
\begin{equation}
m_{X}=\left(\prod_{f:\mu_{1}<m_{f}<\mu_{0}}m_{f}^{b^{f}_{X}}\cdot\prod_{s:\mu_{1}<m_{s}<\mu_{0}}m_{s}^{b^{s}_{X}}\right)^{\frac{1}{b^{\mu_{1}}_{X}-b^{\mu_{0}}_{X}}}
\label{treshnaGmX}
\end{equation}
$m_{5}$ can be replaced by a "jump" $s_{5}$ of the $\alpha^{-1}_{5}$ coupling constant at the $m_{X}$
scale
\begin{equation}
s_{5} = \frac{b^{\mu_{0}}_{5}-b^{\mu_{1}}_{5}}{2\pi}\ln\left(\frac{m_{5}}{m_{X}}\right)
\label{jump5}
\end{equation}
Alternatively, $m_{X}$ can be replaced by a "jump" $s_{X}$ of the $\alpha^{-1}_{X}$ coupling constant at the $m_{5}$
scale
\begin{equation}
s_{X} = \frac{b^{\mu_{0}}_{X}-b^{\mu_{1}}_{X}}{2\pi}\ln\left(\frac{m_{X}}{m_{5}}\right)
\label{jumpX}
\end{equation}
Such a replacement makes no difference for analytic 1-loop running but it can simplify the 1,5-loop running considered in section \ref{Sec1,5loop}.

Knowing the field content of the Case I (tables \ref{tab2} and \ref{tab3} in Appendix \ref{ApTab}), one can calculate all values of $b$ parameters that are present in gauge RGEs. They are shown in Table \ref{tabI}.

\begin{table*}
\caption{Values of $b$ parameters in Cases IA and IB.}
\begin{math}\begin{array}{||c|c|c|c|c|c||c||}
\hline \hline \multicolumn{6}{||c||}{\text{Groups and } b \text{ parameters}}                & \text{Scales}  \\
\hline\hline \multicolumn{6}{||c||}{\textcolor{white}{\stackrel{I}{I}} b^{M_{Pl}}_{10}=\frac{121}{2}} &  M_{Pl}  \\
\hline \multicolumn{6}{||c||}{\textcolor{violet}{SO(10)}}                     &  \textcolor{violet}{\mu_{0}}- M_{Pl} \\
\hline \multicolumn{6}{||c||}{\textcolor{white}{\stackrel{I}{I}} b^{\mu_{0}}_{10}=\frac{121}{2}} &  \textcolor{violet}{\mu_{0}}  \\
\cline{1-6} \multicolumn{4}{||c|}{\textcolor{white}{\stackrel{I}{I}} b^{\mu_{0}}_{5}=127} & \multicolumn{2}{c||}{b^{\mu_{0}}_{X}=137}&\\
\hline \multicolumn{4}{||c|}{\textcolor{blue}{SU(5)}} &  \multicolumn{2}{c||}{U(1)_{X}} & \textcolor{blue}{\mu_{1}} - \textcolor{violet}{\mu_{0}} \\
\hline \multicolumn{4}{||c|}{\textcolor{white}{\stackrel{I}{I}} b^{\mu_{1}}_{5}=-3} &  \multicolumn{2}{c||}{b^{\mu_{1}}_{X}=12}& \\
\cline{1-6} & & \multicolumn{2}{c|}{\textcolor{white}{.......................}}&  \multicolumn{2}{c||}{}& \textcolor{blue}{\mu_{1}} \\
\cline{4-5} \textcolor{white}{\stackrel{I}{I}} b^{\mu_{1}}_{3}=-3&  b^{\mu_{1}}_{2}=1& b^{\mu_{1}}_{\widehat{1} \widehat{1}}=\frac{33}{5} &  \multicolumn{2}{c|}{b^{\mu_{1}}_{\widehat{1} \widehat{2}} =-\frac{\sqrt{6}}{5}} & b^{\mu_{1}}_{\widehat{2} \widehat{2}}= \frac{57}{5}  &     \\
\cline{3-6} \textcolor{white}{\stackrel{I}{I}}  &   &  \multicolumn{4}{c||}{b^{\mu_{1}}_{\underline{1} \underline{1}}=11 \text{, }  b^{\mu_{1}}_{\underline{1} \underline{2}} =8\text{, }  b^{\mu_{1}}_{\underline{2} \underline{2}}=24}&   \\
\hline SU(3)_{c} &  SU(2)_{L} & \multicolumn{4}{c||}{\textcolor{red}{U(1)^{2}}} & \textcolor{red}{M_{Z'}} - \textcolor{blue}{\mu_{1}} \\
\hline \textcolor{white}{\stackrel{I}{I}} b^{M_{Z'}}_{3}=-3 & b^{M_{Z'}}_{2}=1 & \multicolumn{4}{c||}{b^{M_{Z'}}_{\widehat{1} \widehat{1}}=\frac{33}{5}\text{, }b^{M_{Z'}}_{\widehat{1} \widehat{2}} =-\frac{\sqrt{6}}{5}\text{, }b^{M_{Z'}}_{\widehat{2} \widehat{2}}= \frac{57}{5}}   &   \text{Case IB}  \\
\cline{3-6} \textcolor{white}{\stackrel{I}{I}}  &   &  \multicolumn{4}{c||}{b^{M_{Z'}}_{Y}=b^{M_{Z'}}_{\underline{1} \underline{1}}=11 \text{, }  b^{M_{Z'}}_{\underline{1} \underline{2}} =8\text{, }  b^{M_{Z'}}_{\underline{2} \underline{2}}=24}&  \textcolor{red}{M_{Z'}}>T_{SUSY} \\
\hline \textcolor{white}{\stackrel{I}{I}} b^{M_{Z'}}_{3}=-7 & b^{M_{Z'}}_{2}=-\frac{19}{6} & \multicolumn{4}{c||}{b^{M_{Z'}}_{\widehat{1} \widehat{1}}=\frac{41}{10}\text{, }b^{M_{Z'}}_{\widehat{1} \widehat{2}} =-\frac{\sqrt{6}}{30}\text{, }b^{M_{Z'}}_{\widehat{2} \widehat{2}}=\frac{169}{60}}   &   \text{Case IA}  \\
\cline{3-6} \textcolor{white}{\stackrel{I}{I}}  &   &  \multicolumn{4}{c||}{b^{M_{Z'}}_{Y}=b^{M_{Z'}}_{\underline{1} \underline{1}}=\frac{41}{6} \text{, }  b^{M_{Z'}}_{\underline{1} \underline{2}} =\frac{16}{3}\text{, }  b^{M_{Z'}}_{\underline{2} \underline{2}}=\frac{26}{3}}&  \textcolor{red}{M_{Z'}}<T_{SUSY} \\
\hline SU(3)_{c} &  \textcolor{brown}{SU(2)_{L}} & \multicolumn{4}{c||}{\textcolor{brown}{U(1)_{Y}}} &\textcolor{brown}{M_{Z}} - \textcolor{red}{M_{Z'}} \\
\hline \textcolor{white}{\stackrel{I}{I}} b^{M_{Z}}_{3}=-7 & b^{M_{Z}}_{2}=-\frac{19}{6} & \multicolumn{4}{c||}{b^{M_{Z}}_{Y}=\frac{41}{6}}   & \textcolor{brown}{M_{Z}} \\
\hline\hline
\end{array}\end{math}
\label{tabI}
\end{table*}
Another necessary information is the form of $L_{\widehat{a}\underline{b}}$ transformation, already given in eq. (\ref{case1})

\begin{equation}
L_{\widehat{a}\underline{b}} = \left[\begin{array}{cc}
\frac{\sqrt{15}}{5} & 0 \\
-\frac{\sqrt{10}}{5} & \frac{\sqrt{10}}{4}
\end{array}\right] \label{case1rep}\end{equation}

\begin{figure*}
\resizebox{1\textwidth}{!}{
    \includegraphics{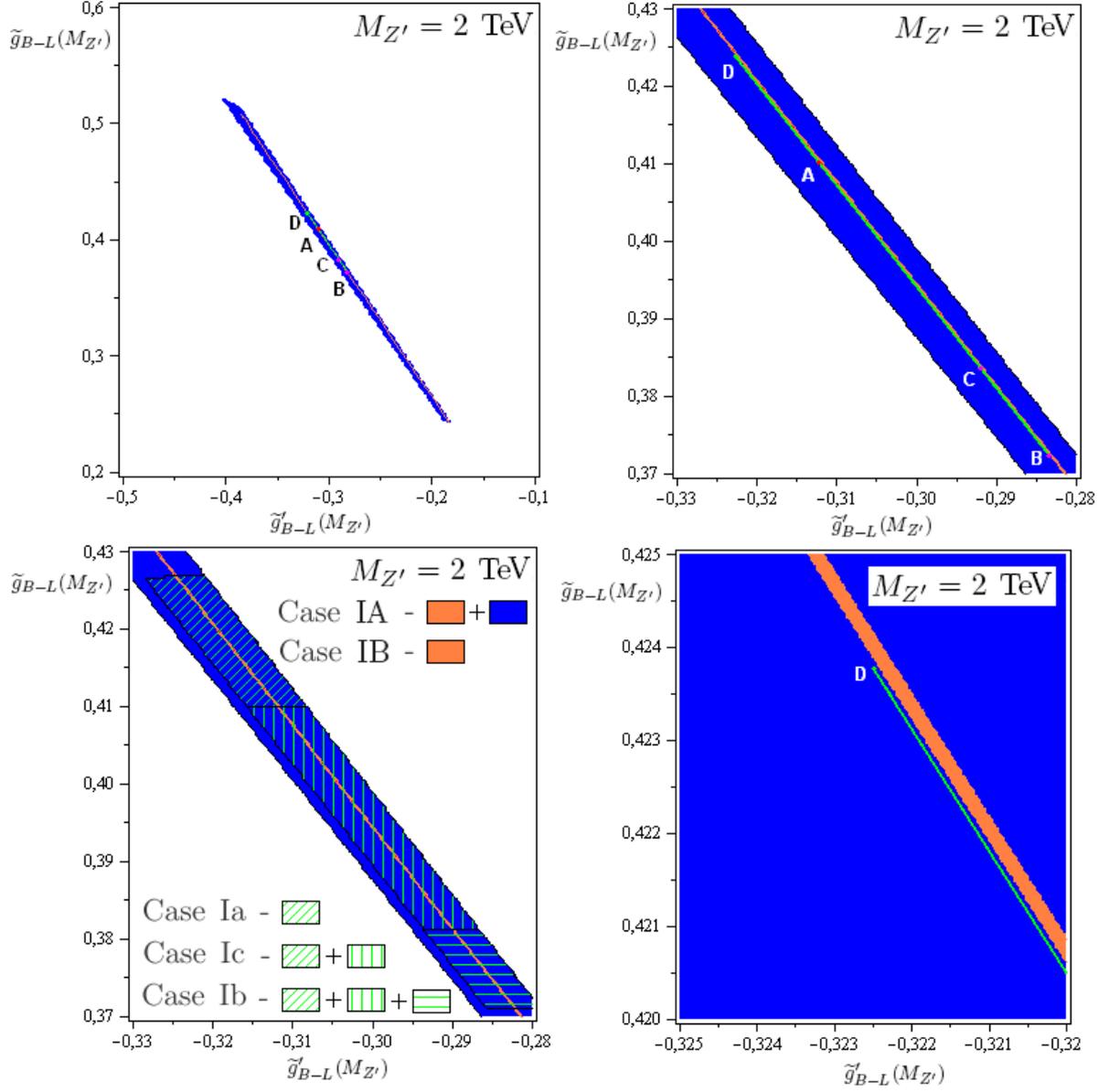}}
  \caption{Regions allowed by 1-loop unification in Case I for $M_{Z'}=2$ TeV. Blue region is allowed only in Case IA and the orange region is allowed in both IA and IB Cases. Green interval with four points (A, B, C and D) is the region allowed by (approximate) unification in simple decoupling pattern with deviations being as small as possible. The lower left plot illustrates the comparison between Cases IA and IB (threshold corrections; $D_{1}=D_{2}=D_{3}=0$; values of $g_{2}(M_{Z})$, $g_{3}(M_{Z})$ and $g'(M_{Z})$ exactly equal to central experimental ones) and Cases Ia, Ib and Ic in simple decoupling pattern (no threshold corrections; $D_{1}$, $D_{2}$, $D_{3}$ smaller than $0.02$; values of $g_{2}(M_{Z})$, $g_{3}(M_{Z})$ and $g'(M_{Z})$ differ from central experimental ones no more than $2\sigma$). All regions for simple decoupling pattern are already shown and described in more detail on Figure \ref{figI}. The upper left plot shows the largest region of parameter space spanned by $\widetilde{g}'_{B-L}(M_{Z'})$ and $\widetilde{g}'_{B-L}(M_{Z'})$ parameters. Other plots show its zoomed parts. Analogous plots for $M_{Z'}$ equal to $2.25$ TeV and $2.5$ TeV look very similar.}
  \label{figSL}
\end{figure*}

We compare the obtained results with the regions allowed in the simple decoupling pattern in Figure \ref{figSL}.

\subsubsection{Case II: Breaking \texorpdfstring{$SO(10) \rightarrow SU(3)_{c} \oplus SU(2)_{L} \oplus  SU(2)_{R} \oplus U(1)_{\widehat{B-L}}$}{\space}}{\label{SubSubSecTreshII}}

There are four gauge coupling constants related to the intermediate group $SU(3)_{c} \oplus SU(2)_{L} \oplus  SU(2)_{R} \oplus U(1)_{\widehat{B-L}}$ - $g_{3}$,  $g_{2}$, $g_{2R}$ and $g_{\widehat{B-L}}$ respectively. $m_{3}$, $m_{2}$, $m_{2R}$ and $m_{\widehat{B-L}}$ are related mass parameters of effective threshold corrections caused by "heavy" fields.
Since $\mu_{1}$ can be as low as $10$ TeV and "heavy" fields have mass terms with mass parameters proportional to $\mu_{0}$, we additionally assume that all of them are heavier than $10^{-3} \mu_{0}$. Then eq. (\ref{treshnaG}) provides the same limit for all four effective threshold mass parameters.

Any three effective mass parameters can be replaced by "jumps" of related inverse coupling constants ($\alpha^{-1}$) at the scale equal to the value of the fourth effective mass parameter.

Knowing the field content of the Case II (tables \ref{tab5}, \ref{tab8} and \ref{tab6} in Appendix \ref{ApTab}), one can calculate all values of $b$ parameters, that are present in gauge RGEs. They are shown in Table \ref{tabII}.

\begin{figure*}
\centering
\resizebox{0.5\textwidth}{!}{
    \includegraphics{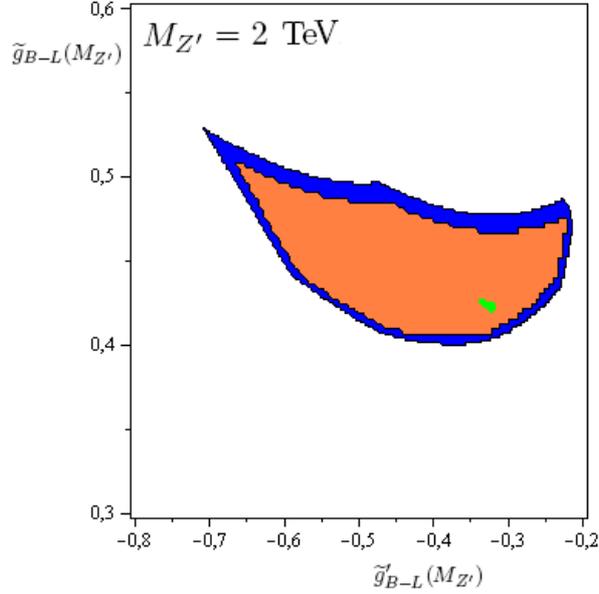}}
  \caption{Regions allowed by 1-loop unification in Case II for $M_{Z'}=2$ TeV. Blue region is allowed only in Case IIA and the orange one is allowed in both IIA and IIB Cases. Small green region is the sum of three regions that are allowed by unification in simple decoupling pattern in Cases IIa, IIb and IIc for deviations $D'_{1}$, $D'_{2}$, $D'_{3}$ smaller than $0.02$ for values of $g_{2}(M_{Z})$, $g_{3}(M_{Z})$ and $g'(M_{Z})$ that differ from central experimental ones no more than $2\sigma$. These regions are already shown and described in more detail on Figure \ref{figII}. Analogous plots for $M_{Z'}$ equal to $2.25$ TeV and $2.5$ TeV look very similar.}
\label{fig7}
\end{figure*}

\begin{table*}
\caption{Values of $b$ parameters in Cases IIA and IIB.}
\begin{math}\begin{array}{||c|c|c|c|c|c||c||}
\hline \hline \multicolumn{6}{||c||}{\text{Groups and } b \text{ parameters}}                & \text{Scales}  \\
\hline\hline \multicolumn{6}{||c||}{ \textcolor{white}{\stackrel{I}{I}} b^{M_{Pl}}_{10}=\frac{69}{2}} &  M_{Pl}  \\
\hline \multicolumn{6}{||c||}{\textcolor{violet}{SO(10)}}                     &  \textcolor{violet}{\mu_{0}}- M_{Pl} \\
\hline \multicolumn{6}{||c||}{ \textcolor{white}{\stackrel{I}{I}} b^{\mu_{0}}_{10}=\frac{69}{2}} &  \textcolor{violet}{\mu_{0}}  \\
\cline{1-6} \textcolor{white}{\stackrel{I}{I}} b^{\mu_{0}}_{3}=76 & b^{\mu_{0}}_{2}=81 & \multicolumn{2}{c|}{b^{\mu_{0}}_{2\,R}=81} & \multicolumn{2}{c||}{b^{\mu_{0}}_{B-L}=82}&\\
\hline SU(3)_{c} &  SU(2)_{L} & \multicolumn{2}{c|}{\textcolor{blue}{SU(2)_{R}}} &  \multicolumn{2}{c||}{U(1)_{B-L}} & \textcolor{blue}{\mu_{1}} - \textcolor{violet}{\mu_{0}} \\
\hline \textcolor{white}{\stackrel{I}{I}} & & \multicolumn{2}{c|}{b^{\mu_{1}}_{2\,R}=7} &  \multicolumn{2}{c||}{b^{\mu_{1}}_{B-L}=15}& \\
\cline{3-6}  b^{\mu_{1}}_{3}=-3 &  b^{\mu_{1}}_{2}=1  & \multicolumn{2}{c|}{\textcolor{white}{.......................}}&  \multicolumn{2}{c||}{}&  \textcolor{blue}{\mu_{1}} \\
\cline{4-5} \textcolor{white}{\stackrel{I}{I}} & & b^{\mu_{1}}_{\widehat{1} \widehat{1}}= 9 &  \multicolumn{2}{c|}{b^{\mu_{1}}_{\widehat{1} \widehat{2}} =-\sqrt{6}} & b^{\mu_{1}}_{\widehat{2} \widehat{2}}= 9  &     \\
\cline{3-6} \textcolor{white}{\stackrel{I}{I}}  &   &  \multicolumn{4}{c||}{b^{\mu_{1}}_{\underline{1} \underline{1}}=11 \text{, }  b^{\mu_{1}}_{\underline{1} \underline{2}} =8\text{, }  b^{\mu_{1}}_{\underline{2} \underline{2}}=24}&   \\
\hline SU(3)_{c} &  SU(2)_{L} & \multicolumn{4}{c||}{\textcolor{red}{U(1)^{2}}} & \textcolor{red}{M_{Z'}} - \textcolor{blue}{\mu_{1}} \\
\hline \textcolor{white}{\stackrel{I}{I}} b^{M_{Z'}}_{3}=-3 & b^{M_{Z'}}_{2}=1 & \multicolumn{4}{c||}{b^{M_{Z'}}_{\widehat{1} \widehat{1}}=9\text{, }b^{M_{Z'}}_{\widehat{1} \widehat{2}} =-\sqrt{6}\text{, }b^{M_{Z'}}_{\widehat{2} \widehat{2}}= 9}   &   \text{Case IIB}  \\
\cline{3-6} \textcolor{white}{\stackrel{I}{I}}  &   &  \multicolumn{4}{c||}{b^{M_{Z'}}_{Y}=b^{M_{Z'}}_{\underline{1} \underline{1}}=11 \text{, }  b^{M_{Z'}}_{\underline{1} \underline{2}} =8\text{, }  b^{M_{Z'}}_{\underline{2} \underline{2}}=24}&  \textcolor{red}{M_{Z'}}>T_{SUSY} \\
\hline \textcolor{white}{\stackrel{I}{I}} b^{M_{Z'}}_{3}=-7 & b^{M_{Z'}}_{2}=-\frac{19}{6} & \multicolumn{4}{c||}{b^{M_{Z'}}_{\widehat{1} \widehat{1}}=\frac{11}{3}\text{, }b^{M_{Z'}}_{\widehat{1} \widehat{2}} =\frac{\sqrt{6}}{4}\text{, }b^{M_{Z'}}_{\widehat{2} \widehat{2}}=\frac{13}{4}}   &   \text{Case IIA}  \\
\cline{3-6} \textcolor{white}{\stackrel{I}{I}}  &   &  \multicolumn{4}{c||}{b^{M_{Z'}}_{Y}=b^{M_{Z'}}_{\underline{1} \underline{1}}=\frac{41}{6} \text{, }  b^{M_{Z'}}_{\underline{1} \underline{2}} =\frac{16}{3}\text{, }  b^{M_{Z'}}_{\underline{2} \underline{2}}=\frac{26}{3}}&  \textcolor{red}{M_{Z'}}<T_{SUSY} \\
\hline SU(3)_{c}  &  \textcolor{brown}{SU(2)_{L}} & \multicolumn{4}{c||}{\textcolor{brown}{U(1)_{Y}}} &\textcolor{brown}{M_{Z}} - \textcolor{red}{M_{Z'}} \\
\hline \textcolor{white}{\stackrel{I}{I}} b^{M_{Z}}_{3}=-7 & b^{M_{Z}}_{2}=-\frac{19}{6} & \multicolumn{4}{c||}{b^{M_{Z}}_{Y}=\frac{41}{6}}   & \textcolor{brown}{M_{Z}} \\
\hline\hline
\end{array}\end{math}
\label{tabII}
\end{table*}
To avoid rapid proton decay we assume that all six $SU(3)_{c}$ triplets that can mediate it have masses equal to $\mu_{0}$ (as in Cases IIb and IIc). Another necessary information is the form of $L_{\widehat{a}\underline{b}}$ transformation, already given in eq. (\ref{case2})

\begin{equation}
L_{\widehat{a}\underline{b}} = \left[\begin{array}{cc}
1 & -\frac{1}{2} \\
0 & \frac{\sqrt{6}}{4}
\end{array}\right] \label{case2rep}\end{equation}
In Figure \ref{fig7} we compare obtained results with the sum of regions allowed in simple decoupling pattern.
These results are identical to the ones obtained without additional $10^{-3} \mu_{0}$ bound. This fact could be explained by the lost of perturbativity that happens often when $m_{3}$, $m_{2}$, $m_{2R}$ or $m_{\widehat{B-L}}$ is smaller than $10^{-3} \mu_{0}$.

\section{1,5-loop running of gauge and Yukawa coupling constants in Case IB}{\label{Sec1,5loop}}

The 1,5 loop running is based on 2-loop RGE equations for gauge coupling constants and 1-loop RGEs for Yukawa coupling constants. Details are shown in Appendix \ref{ApRGE}. We performed such a bottom-up running numerically in the $X_{\widehat{a}}$ basis for Case IB for $M_{Z'} = 2.5$ TeV. The running always started at $M_{Z}$ scale. W've selected six points - $P_{1}$, $P_{2}$, $\ldots P_{6}$ in the full parameter space, that give exact 1-loop unification to $SU(5)$ at the $\mu_{1}$ scale and exact 1-loop unification to $SO(10)$ at the $\mu_{0}$ scale. Values of parameters for these points are shown in Table \ref{tabP1}. Each point is uniquely defined by values of seven parameters $\widetilde{g}_{B-L}(M_{Z'})$, $\widetilde{g}'_{B-L}(M_{Z'})$, $T_{SUSY}$, $s_{2}$, $s_{3}$, $\ln(\frac{m_{X}}{\mu_{1}})$ and $\ln(\frac{m_{5}}{\mu_{1}})$. We recall that $\widetilde{g}_{B-L}(M_{Z'})$ and $\widetilde{g}'_{B-L}(M_{Z'})$ are two gauge coupling constants, normalized as shown in formula (\ref{Zwircon}). $T_{SUSY}$ is one of effective threshold mass parameters, defined by formula (\ref{treshaTSUSY}). "jumps" $s_{2}$ and $s_{3}$ are defined by formula (\ref{jumps}). $m_{5}$ and $m_{X}$ are effective threshold mass parameters defined by formulas (\ref{treshnaGm5}) and (\ref{treshnaGmX}) respectively.

\begin{table}
\caption{Values of parameters for six points that correspond to exact 1-loop unification in Case IB. $\mu_{1}$ and $\mu_{0}$ are not independent parameters, since their values are determined by the 1-loop running.}
\begin{math}\begin{array}{|c|r|r|r|r|r|r|}
\hline\hline \text{Points}&\multicolumn{1}{c|}{P_{1}}&\multicolumn{1}{c|}{P_{2}}&\multicolumn{1}{c|}{P_{3}}&\multicolumn{1}{c|}{P_{4}}&\multicolumn{1}{c|}{P_{5}}& \multicolumn{1}{c|}{P_{6}}\\
\hline\hline \widetilde{g}_{B-L}(M_{Z'})  &  0.500 &  0.450 &  0.400 &  0.350 &  0.300 &  0.250 \\
\hline \widetilde{g}'_{B-L}(M_{Z'})       & -0.380 & -0.342 & -0.304 & -0.266 & -0.228 & -0.190 \\
\hline T_{SUSY} [\text{GeV}]              & 1500     & 1750     & 1750     & 2500     & 2500     & 2000     \\
\hline s_{2}                              & -0.107 & -0.142 & -0.142 & -0.199 & -0.199 & -0.181 \\
\hline s_{3}                              &  0.505 &  0.478 &  0.478 &  0.457 &  0.457 &  0.440 \\
\hline \ln(\frac{m_{X}}{\mu_{1}})         &  1.201 &  0.498 &  3.554 &  1.421 &  2.488 &  0       \\
\hline \ln(\frac{m_{5}}{\mu_{1}})         &  0       &  0       &  3.554 &  2.843 &  5.686 &  6.638 \\
\hline\hline \mu_{1} [10^{16}\text{ GeV}] &  1.050 &  1.043 &  1.043 &  1.000 &  1.000 &  1.047 \\
\hline \mu_{0} [10^{16}\text{ GeV}]       &  3.624 &  2.286 & 93.84  & 60.23  & 883.1  & 1134  \\
\hline\hline 
\end{array}\end{math}
\label{tabP1}
\end{table}

The 1,5-loop running for all these points lead to the following deviations:
\begin{equation} 
\begin{array}{l} D_{1}=\frac{(\vartheta^{-1})_{\widehat{1} \widehat{2}}(\mu_{1})}{(\vartheta^{-1})_{\widehat{1} \widehat{1}}(\mu_{1})}= -\frac{\vartheta_{\widehat{1} \widehat{2}}(\mu_{1})}{\vartheta_{\widehat{2} \widehat{2}}(\mu_{1})}\hspace{0.5cm}
D_{2}=\frac{\alpha_{3}(\mu_{1})-\alpha_{2}(\mu_{1})}{\alpha_{3}(\mu_{1})+\alpha_{2}(\mu_{1})}
\\ D_{3}=\frac{\alpha_{5}(\mu_{0})-\alpha_{X}(\mu_{0})}{\alpha_{5}(\mu_{0})+\alpha_{X}(\mu_{0})} \hspace{0.5cm} D_{4}=\frac{\alpha_{3}(\mu_{1})-\vartheta_{\widehat{1} \widehat{1}}(\mu_{1})}{\alpha_{3}(\mu_{1})+\vartheta_{\widehat{1} \widehat{1}}(\mu_{1})} \end{array}
\end{equation}
For points $P_{1}$, $P_{2}$, $\ldots P_{6}$ these deviations are defined for $\mu_{1}$ and $\mu_{0}$ taken from the 1-loop running. Therefore, we need to introduce the $D_{4}$ deviation. For 1-loop running in simple decoupling pattern (subsection \ref{SubSubSecNotreshI}) $D_{4} = D_{2}$. Here, values of $D_{4}$ can be different than values of $D_{2}$ due to different definition of the $\mu_{1}$ scale.

Apart from $g'_{B-L}(M_{Z'})$, $g_{B-L}(M_{Z'})$, threshold scales and "jumps" there are some other parameters that have to be specified to perform our 1,5-loop bottom-up running. $\tan{\beta}$ and renormalized quark and lepton masses at the $M_{Z}$-scale are needed to calculate initial values of three standard, diagonal, 3rd-generation Yukawa coupling constants - $Y_{t}$, $Y_{b}$ and $Y_{\tau}$ (we neglect other parts of $3\times3$ Yukawa matrices). We put $\tan{\beta}=10$ and get needed masses at the $M_{Z}$-scale from \cite{M_Z masses}. Moreover, in the Interval $I_{M_{Z}}^{\mu_{1}}$, we consider one more Yukawa coupling constant - $Y_{R}$. We assume that it starts to run up from the $M_{Z'}$ scale, because this scale is close to the mass of $\nu_{R}$. When $\nu_{R}$ is integrated out below its mass, its Yukawa couplings with $Y_{R}$ and $Y_{\nu}$ are naturally eliminated from the effective theory. We put $Y_{R}(M_{Z'})=0.9$ and neglect $Y_{\nu}$ because it should be very small ($Y_{\nu}(M_{Z'}) \sim 10^{-5}$) to obtain right order of magnitude for light neutrino masses. All other Yukawa coupling constants, which could be present in the Interval $I_{M_{Z}}^{\mu_{1}}$, are related to fields that are integrated out in this Interval. $Y_{t}$, $Y_{b}$, $Y_{\tau}$ and $Y_{\nu}$ originate from mixings between $Y_{10}$ and $Y_{\overline{126}}$. As mention in section \ref{SecSO10}, we don't consider details of these mixings, so we don't demand any form of $b-\tau$ Yukawa unification at $\mu_{1}$ scale. In the Interval $I_{\mu_{1}}^{\mu_{0}}$ we consider only $Y_{R}$ and two Yukawa coupling constants that originate from $Y_{10}$. Their initial values at the $\mu_{1}$ scale are determined by $Y_{t}$ and $Y_{b}$ neglecting all Yukawa coupling constants that originate from $Y_{\overline{126}}$ together with all eventual mixings. That is the reason why we don't demand any Yukawa unification at the $\mu_{0}$ scale. Yukawa RGEs and Yukawa terms in gauge RGEs are always supersymmetric even near the $M_{Z}$ scale. It's a convenient approximation that allows to avoid very complicated equations with 4-scalar coupling constants (Appendix \ref{ApRGE}).

Deviations $D_{1}$, $D_{2}$, $D_{3}$ and $D_{4}$ for points $P_{1}$, $P_{2}$, $\ldots P_{6}$ for the 1,5-loop running are shown in Table \ref{tabP2}

\begin{table*}
\caption{Deviations $D_{1}$, $D_{2}$, $D_{3}$ and $D_{4}$ for six points from Table \ref{tabP1} for the 1,5-loop running. For the point $P_{1}$ the $D_{3}$ deviation cannot be determined since $\alpha_{X}$ looses perturbativity at the scale $2.90399\cdot10^{16}$ GeV, which is smaller than the $\mu_{0}$ scale - $3.62447\cdot10^{16}$ GeV.}
\begin{math}\begin{array}{|c|r|r|r|r|r|r|}
\hline\hline \text{Points}&\multicolumn{1}{c|}{P_{1}}&\multicolumn{1}{c|}{P_{2}}&\multicolumn{1}{c|}{P_{3}}&\multicolumn{1}{c|}{P_{4}}&\multicolumn{1}{c|}{P_{5}}& \multicolumn{1}{c|}{P_{6}}\\
\hline\hline D_{1}                  & 0.00283    & 0.00270  & 0.00267    & 0.00251    & 0.00250    & 0.00259 \\
\hline D_{2}                        & -0.01003   & -0.00993 & -0.00989   & -0.00985   & -0.00983   & -0.00986 \\
\hline D_{3}                        & -          & -0.254   &  0.0362    &  0.211     &  0.0749    &  0.702     \\
\hline D_{4}                        & -0.00355   & -0.00354 & -0.00351   & -0.00366   & -0.00365   & -0.00355 \\
\hline\hline 
\end{array}\end{math}
\label{tabP2}
\end{table*}

To cancel deviations $D_{1}$, $D_{2}$, $D_{3}$ and $D_{4}$ one can try to slightly change some parameters of points $P_{1}$, $P_{2}$, $\ldots P_{6}$. Corrected points are denoted by $P_{1}'$, $P_{2}'$, $\ldots P_{6}'$. We'd like to change only $\widetilde{g}_{B-L}(M_{Z'})$ and $\widetilde{g}'_{B-L}(M_{Z'})$ since we are interested in the influence of the 1,5-loop effects on these two additional gauge coupling constants. Then, one can cancel $D_{1}$ and $D_{3}$. To cancel $D_{2}$ we additionally change the $s_{3}$ jump. In practice we demand all three deviations to be smaller than $10^{-5}$ and for each point we find new values of $\widetilde{g}_{B-L}(M_{Z'})$, $\widetilde{g}'_{B-L}(M_{Z'})$ and $s_{3}$ that differ from original ones as little as possible. The scales $\mu_{0}$ and $\mu_{1}$ are now set by the 1,5-loop running (and not taken from the corresponding 1-loop running). Since the unification is only very close to the exact one, the $\mu_{1}$ scale is formally defined as the scale at which $\alpha_{2} = \vartheta_{\widehat{1} \widehat{1}}$. Therefore $D_{4}$ is just equal to $D_{2}$ as in simple decoupling pattern. Results are shown in Table \ref{tabP3}.

\begin{table*}
\caption{Points $P_{1}$, $P_{2}$, $\ldots P_{6}$ compared to modified points $P_{1}'$, $P_{2}'$, $\ldots P_{6}'$ that correspond to almost exact 1,5-loop unification in Case IB. Values of $\mu_{1}$ and $\mu_{0}$ are determined by the 1,5-loop running. $\Delta$ denotes differences between values of parameters for new points and values of parameters for old points.}
\begin{math}\begin{array}{|c|r|r|r|r|r|r|}
\hline\hline\text{Old Points}&\multicolumn{1}{c|}{P_{1}}&\multicolumn{1}{c|}{P_{2}}&\multicolumn{1}{c|}{P_{3}}&\multicolumn{1}{c|}{P_{4}}&\multicolumn{1}{c|}{P_{5}}& \multicolumn{1}{c|}{P_{6}}\\
\hline \widetilde{g}_{B-L}(M_{Z'})  &  0.500   &  0.450   &  0.400   &  0.350   &  0.300   &  0.250   \\
\hline \widetilde{g}'_{B-L}(M_{Z'}) & -0.380   & -0.342   & -0.304   & -0.266   & -0.228   & -0.190  \\
\hline s_{3}                        &  0.505   &  0.478   &  0.478   &  0.457   &  0.457   &  0.440   \\
\hline\hline \text{New Points}&\multicolumn{1}{c|}{P_{1}'}&\multicolumn{1}{c|}{P_{2}'}&\multicolumn{1}{c|}{P_{3}'}&\multicolumn{1}{c|}{P_{4}'}&\multicolumn{1}{c|}{P_{5}'}& \multicolumn{1}{c|}{P_{6}'}\\
\hline \widetilde{g}_{B-L}(M_{Z'})  &  0.494   &  0.450   &  0.402   &  0.352   &  0.302   &  0.250   \\
\hline \widetilde{g}'_{B-L}(M_{Z'}) & -0.375   & -0.342   & -0.305   & -0.268   & -0.229   & -0.190   \\
\hline s_{3}                        & -0.258   & -0.292   & -0.289   & -0.305   & -0.304   & -0.324   \\
\hline D_{1}                        & <10^{-5}   & <10^{-5}   & <10^{-5}   & <10^{-5}   & <10^{-5}   & <10^{-5}   \\
\hline D_{2}=D_{4}                  & -0.00025   & <10^{-5}   & <10^{-5}   & <10^{-5}   & <10^{-5}   & <10^{-5}   \\
\hline D_{3}                        & <10^{-5}   &  0         &  0         &  0         &  0         & <10^{-5}   \\
\hline \mu_{1} [10^{16}\text{ GeV}] & 1.544    & 1.525    &  1.524   &  1.448   &   1.447  &    1.525 \\
\hline \mu_{0} [10^{16}\text{ GeV}] & 5.153    & 3.369    & 80.87    & 36.87    & 677.8    & 1221    \\
\hline\hline \text{Shifts}&\multicolumn{1}{c|}{P_{1}'-P_{1}}&\multicolumn{1}{c|}{P_{2}'-P_{2}}&\multicolumn{1}{c|}{P_{3}'-P_{3}}&\multicolumn{1}{c|}{P_{4}'-P_{4}}&\multicolumn{1}{c|}{P_{5}'-P_{5}}& \multicolumn{1}{c|}{P_{6}'-P_{6}}\\
\hline \Delta\widetilde{g}_{B-L}(M_{Z'})  & -0.00576   & 0         &  0.00220   &  0.00246   &  0.00151   &  0.00039   \\
\hline \Delta\widetilde{g}'_{B-L}(M_{Z'}) &  0.00535   & 0.00086   & -0.00092   & -0.00124   & -0.00061   & 0.00017   \\
\hline \Delta s_{3}                       & -0.763   & -0.770   & -0.767  & -0.762   & -0.761   & -0.764   \\
\hline\hline
\end{array}\end{math}
\label{tabP3}
\end{table*}

Unfortunately for the point $P_{1}'$ we have $|D_{2}| = 2.5\cdot 10^{-4} > 10^{-5}$. It's because values of $s_{3}$ that could make this deviation smaller are forbidden by the $M_{3}>1$ TeV inequality.  

1,5-loop results are generally similar to 1-loop ones with one minor exception. 2-loop terms in gauge RGEs are usually positive, so they cause gauge coupling constants to grow faster, than in 1-loop running. The faster a gauge coupling constant grows, the earlier it looses perturbativity. Therefore, perturbativity constraints are stronger at 1,5-loop level. That is the reason, why the largest values of $g_{B-L}(M_{Z'})$ that are still allowed by 1-loop unification ($\widetilde{g}_{B-L}(M_{Z'}) = 0.50000$ for $P_{1}$ point) are already forbidden by 1,5-loop unification ($\widetilde{g}_{B-L}(M_{Z'})$ reduced to $0.494$ for $P_{1}'$ point). The shift of $\widetilde{g}_{B-L}(M_{Z'})$ between $P_{1}$ and $P_{1}'$ is equal to $-0.00576$. It's the largest shift of this value from all considered ones and the only negative one. Such a shift was necessary to satisfy stronger perturbativity constraints.

\section{Comparison with experimental constraints}{\label{SecDosw}}

\begin{figure*}
\resizebox{1\textwidth}{!}{
\includegraphics{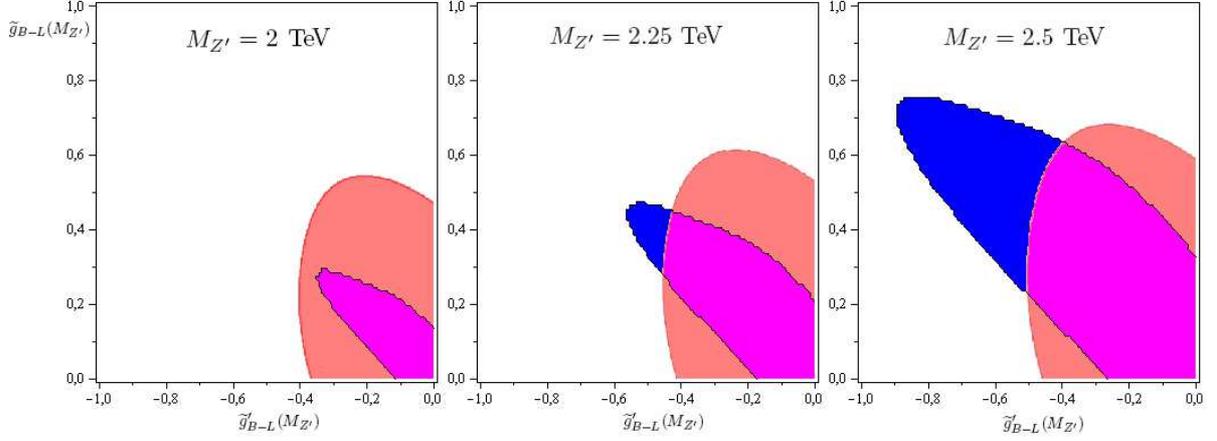}}
\caption{Experimental constraints on $\widetilde{g}'_{B-L}$ and $\widetilde{g}_{B-L}$ for $M_{Z'}$ equal to $2$ TeV (left plot), $2.25$ TeV (middle plot) and $2.5$ TeV (right plot). Blue and magenta regions are allowed by CMS. Dark orange and magenta regions are allowed by EWPT.}
\label{fig8}
\end{figure*}

\begin{figure*}
\resizebox{1\textwidth}{!}{
\includegraphics{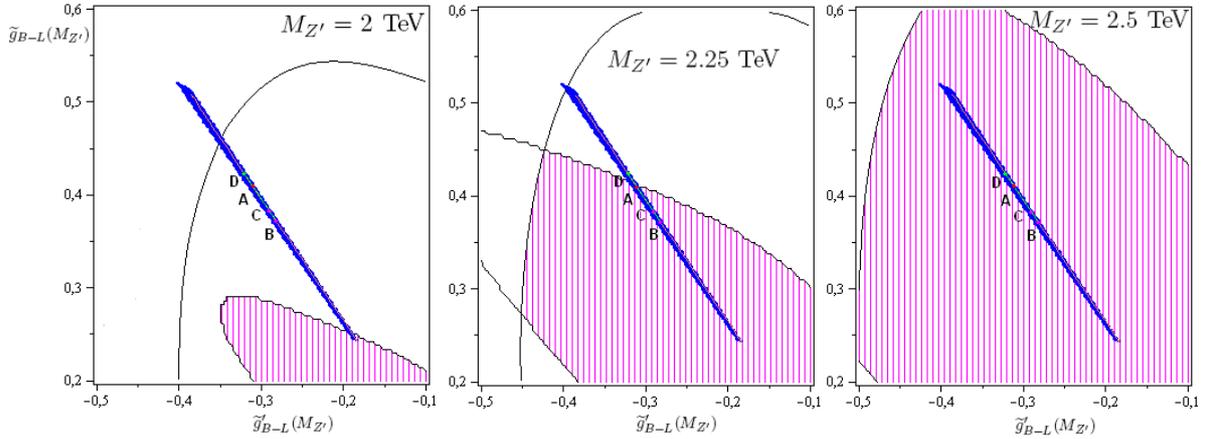}}
\caption{1-loop unification constraints in Case IA (blue and orange regions) and in Case IB (only orange region) compared to experimental ones (magenta stripes) for $\widetilde{g}'_{B-L}$ and $\widetilde{g}_{B-L}$ with $M_{Z'}$ equal to $2$ TeV (left plot), $2.25$ TeV (middle plot) and $2.5$ TeV (right plot). The green interval with four points (A, B, C and D) is allowed by unification in simple decoupling pattern and it's already shown and described on Figure \ref{figI}. We additionally show all lines that correspond to the CMS and EWPT limits. They can be easily identified by comparing this figure to figure \ref{fig8}. The inner region is always the allowed one.}
\label{fig9}
\end{figure*}

\begin{figure*}
\resizebox{1\textwidth}{!}{
\includegraphics{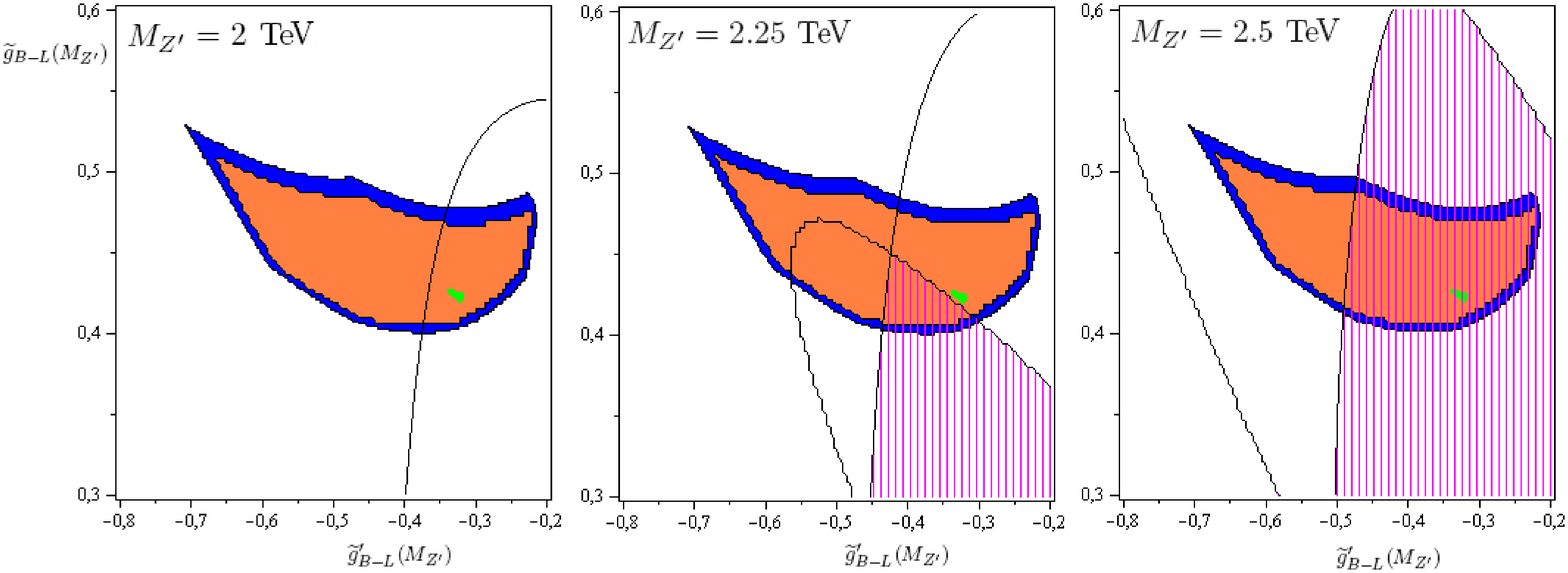}}
\caption{1-loop unification constraints in Case IIA (blue and orange regions) and in Case IIB (only orange region) compared to experimental ones (magenta stripes) for $\widetilde{g}'_{B-L}$ and $\widetilde{g}_{B-L}$ with $M_{Z'}$ equal to $2$ TeV (left plot), $2.25$ TeV (middle plot) and $2.5$ TeV (right plot). The small green region is the sum of regions allowed by unification in simple decoupling pattern. These regions are already shown and described on Figure \ref{figII}. We additionally show all lines that correspond to the CMS and EWPT limits. They can be easily identified by comparing this figure to figure \ref{fig8}. The inner region is always the allowed one.}
\label{fig10}
\end{figure*}

Important experimental constraints on the minimal $Z'$ model are currently provided not only by the LHC, but also by electroweak precision tests (EWPT). They have been shown as constraints on $\widetilde{g}'_{B-L}$ and $\widetilde{g}_{B-L}$ for different values of $M_{Z'}$ in \cite{Zwirner2}.
The crucial LHC data are taken from the $95\%$ C.L. exclusion plot published by CMS collaboration \cite{CMS}. It shows an upper limit for the total cross-section in the $Z'\longrightarrow l^{+}l^{-}$ channel divided by analogous cross-section for the $Z$ boson ($l$ is an electron or a muon). To obtain needed constraints, we calculated this ratio in the LO (leading order) in the narrow width approximation as a function of $\widetilde{g}'_{B-L}$ and $\widetilde{g}_{B-L}$. Details of this kind of calculation are described in \cite{Zwirner2} and in Appendix \ref{ApCMS}. Constraints from ATLAS are easier to be obtained (the total cross-section in the $Z'\longrightarrow l^{+}l^{-}$ channel is given explicitly instead of the ratio) but they are currently weaker (although more recent) since they include only the data from the $8$ TeV LHC run \cite{ATLAS}. Constraints from CMS include smaller amount of data from the $8$ TeV LHC run, but they also include previous data from the $7$ TeV LHC run and the total amount of included data is larger \cite{CMS}. Experimental constraints from CMS and EWPT are shown in Figure \ref{fig8}.
The procedure, described in section \ref{Sec1loop}, provides 1-loop unification constraints on $\widetilde{g}'_{B-L}$ and $\widetilde{g}_{B-L}$ in Cases I and II. Results, compared with experimental constraints, are shown in Figures \ref{fig9} and \ref{fig10} respectively.
\newline
As we can see, for $M_{Z'}=2$ TeV, Case I is almost excluded and Case II is definitely excluded by experiments. On the other hand, for $M_{Z'}=2.5$ TeV experimental constraints are weak enough to allow all the region of Case I and most of the region of Case II. In Case I the lower limit for the $M_{Z'}$ is very close to $2$ TeV and in Case II it lies in the range $2-2.25$ TeV. In simple decoupling patterns in both Cases it's close to $2.25$ TeV. The CMS constraints on $\widetilde{g}'_{B-L}$ and $\widetilde{g}_{B-L}$ depend strongly on the $M_{Z'}$ since the experimental ability to produce the $Z'$ boson decreases significantly when its assumed mass is getting closer and closer to the $\sqrt{s}$.

\section{Conclusions}{\label{SecPods}}

In this paper we considered two $SO(10)$-GUT extensions of the $Z'$ model. In Case I the $SO(10)$ group is initially broken to the $SU(5)\oplus U(1)_{X}$ group and in Case II it's initially broken to the $SU(3)_{c} \oplus SU(2)_{L} \oplus SU(2)_{R} \oplus U(1)_{B-L}$ group. In our approach we allowed for the large hierarchy between two unification scales - $\mu_{0}$ and $\mu_{1}$. We also gave the large freedom for values of effective threshold mass parameters.

Values of two gauge coupling constants from the $Z'$ model - $\widetilde{g}'_{B-L}$ and $\widetilde{g}_{B-L}$ have been analytically constrained by gauge coupling unification and other necessary conditions at 1-loop. Perturbativity constraints are stronger at 2-loop so the allowed region on the plane spanned by $\widetilde{g}'_{B-L}$ and $\widetilde{g}_{B-L}$ is smaller than at 1-loop. We've checked this in Case IB where other differences between 1-loop and 2-loop values of $\widetilde{g}'_{B-L}$ and $\widetilde{g}_{B-L}$ are small.

The comparison between these theoretical constraints and experimental ones (coming from the CMS and\newline EWPT) has been made. It allowed for setting lower limits on $M_{Z'}$ in both cases. In Case I the limit is very close to $2$ TeV and in Case II it lies in the range $2-2.25$ TeV.

\section{Acknowledgments}

The author would like to thank very much prof. Stefan Pokorski for his help - looking over and inspiring his scientific work, correcting his mistakes, encouraging progress, discussing problems and always asking the most important questions. Another person who deserves for thanks is prof. Marek Olechowski who gave some important ideas and read the preliminary version of this paper to suggest some improvements and corrections.
This work has been partially supported by the following grant: UMO-2011/01/M/ST2/02466.

\appendix
\section{RGEs for more than one \texorpdfstring{$U(1)$}{U(1)} gauge group}{\label{ApRGE}}
The following set of Renormalization Group Equations, which has been used for analytical and numerical calculations in this paper, is a generalization of equations presented in \cite{Aguila} and \cite{Jones}. In \cite{Jones} there are only supersymmetric RGEs which are valid for only one $U(1)$ group. They are 2-loop for gauge coupling constants and 1-loop for Yukawa coupling constants. In \cite{Aguila} there are 2-loop RGEs for gauge coupling constants that are valid for arbitrary number of $U(1)$'s. They are presented for both SUSY and non-SUSY case. However, Yukawa coupling constants are neglected. 

Notation is the following. Capital Greek indices\newline $\Phi, \Psi, \Xi \ldots$ denote single chiral superfields which belong to gauge multiplets (They don't denote the whole multiplets). Analogously, small Latin index $f$ denotes a single fermionic (Weyl) field and small Latin index $s$ denotes a single complex scalar field. Capital Latin indices $A, B\ldots$ denote non-abelian gauge groups. For a gauge group $G_{A}$: $g_{A}$ is the gauge coupling constant; $C_{2}(A)$ is the quadratic Casimir operator of the group (quadratic Casimir operator of the adjoint representation of this group); $C_{2}^{A}(x)$ is the quadratic Casimir operator of the irreducible representation that contains the single field $x$; $S_{2}^{A}(x)$ is the Dynkin index of the irreducible representation that contains the single field $x$; $d_{xA}$ is the dimension of the irreducible representation that contains the single field $x$. Lower case Latin indices $a,b,c,d,e$ denote abelian $U(1)$ gauge groups. For groups $U(1)_{a}$ and $U(1)_{b}$: $G_{ab}$ is the gauge coupling constant; $X^{x}_{a}$ and $X^{x}_{b}$ are charges of the single field $x$. $Y^{\Phi \Psi \Xi}$ are Yukawa coupling constants for single fields denoted by $\Phi$, $\Psi$, $\Xi$ and the Yukawa coupling is the following: $\frac{1}{3!}Y^{\Phi \Psi \Xi}\,\Phi\Psi\Xi$.

The general form of the RGE for a coupling constant $g_A$ is the following

\begin{equation}
\mu \frac{d}{d \mu} g_{A} =  \frac{1}{(4\pi)^{2}} g_{A}^{3} b_{A}
\label{gaRGE}
\end{equation}
One can simplify eq. (\ref{gaRGE}) and other ones by replacing $g_{A}$ with $\alpha_{A}=\frac{1}{4\pi}g_{A}^{2}$ (without introducing any square-roots of new variables). Simplified equation has the general form

\begin{equation}
\mu \frac{d}{d \mu} \alpha_{A} =  \frac{1}{2\pi} \alpha_{A}^{2} b_{A}
\label{alfaRGE}
\end{equation}
The general form of the RGE for a coupling constant $G_{ab}$ is the following

\begin{equation}
\mu \frac{d}{d \mu} G_{a b} =  \frac{1}{(4\pi)^{2}} G_{a c} G^{T}_{c d} b_{d e} G_{e b}
\label{GnaRGE}
\end{equation}
For any given loop-order $b_{de}$ is a polynomial in dimensionless coupling constants. For any Feynman diagram every internal abelian propagator is related to expression $(GG^{T})_{ab}$ (provided it's proportional to $\delta_{ab}$). As a consequence, $b_{de}$ (and $\beta$-functions in other RGEs) depends on abelian coupling constants $G_{ab}$ only through the $(GG^{T})_{ab}$ expression (symmetric product). Therefore, one can simplify equation (\ref{GnaRGE}) and other ones by replacing $G_{ab}$ with $\vartheta_{ab}=\frac{1}{4\pi}(GG^{T})_{ab}$ (without introducing any square-roots). Simplified equations have the general form

\begin{equation}
\mu \frac{d}{d \mu} \vartheta_{ab} =  \frac{1}{2\pi} \vartheta_{ad} b_{de} \vartheta_{eb}
\label{AnaRGE}
\end{equation}
$b_{A}$ and $b_{d e}$ parameters in equations (\ref{alfaRGE}) and (\ref{AnaRGE}) contain the following parts:\newline
1-loop:

\begin{equation}
(b_{A})^{1-loop}= -\frac{11}{3} C_{2}(A) + \frac{2}{3}\sum_{f}\frac{S^{A}_{2}(f)}{d_{fA}} + \frac{1}{3}\sum_{s}\frac{S^{A}_{2}(s)}{d_{sA}}
\label{bA1-loop}
\end{equation}

\begin{equation}
(b_{d e})^{1-loop}= \frac{2}{3} \sum_{f}(X^{f}_{d} X^{f}_{e}) + \frac{1}{3}\sum_{s}(X^{s}_{d} X^{s}_{e})
\end{equation}
1-loop - SUSY case:

\begin{equation}
(b_{A})^{1-loop}_{SUSY}= -3 C_{2}(A) + \sum_{\Phi}\frac{S^{A}_{2}(\Phi)}{d_{\Phi A}} 
\end{equation}

\begin{equation}
(b_{d e})^{1-loop}_{SUSY}= \sum_{\Phi}(X^{\Phi}_{d} X^{\Phi}_{e})
\end{equation}
2-loop:

\begin{equation}\begin{array}{l}
(b_{A})^{2-loop}=  -\frac{17}{24\pi^{2}} (C_{2}(A))^{2} \\ + \frac{1}{2\pi} \sum_{f}\left[ \frac{S^{A}_{2}(f)}{d_{f A}} \left(X^{f}_{a}\vartheta_{ab}X^{f}_{b} + \alpha_{B} \left(C^{B}_{2}(f)+\frac{5}{3}\delta^{BA}C_{2}(A)\right)\right)\right] \\ + \frac{1}{\pi} \sum_{s}\left[ \frac{S^{A}_{2}(s)}{d_{s A}} \left(X^{s}_{a}\vartheta_{ab}X^{s}_{b} + \alpha_{B} \left(C^{B}_{2}(s)+\frac{1}{6}\delta^{BA}C_{2}(A)\right)\right)\right] \\ +\,\text{terms with Yukawa and 4-scalar coupling constants}
\end{array}\end{equation}
\begin{equation}\begin{array}{l}
(b_{d e})^{2-loop}=  \frac{1}{2\pi} \sum_{f}\left[X^{f}_{d} X^{f}_{e} \left(X^{f}_{a}\vartheta_{ab}X^{f}_{b} + \alpha_{B} C^{B}_{2}(f)\right)\right] \\+ \frac{1}{\pi} \sum_{s}\left[X^{s}_{d} X^{s}_{e} \left(X^{s}_{a}\vartheta_{ab}X^{s}_{b} + \alpha_{B} C^{B}_{2}(s)\right)\right] \\ +\,\text{terms with Yukawa and 4-scalar coupling constants}
\end{array}\end{equation}
2-loop - SUSY case:

\begin{equation}\begin{array}{l}
(b_{A})^{2-loop}_{SUSY}= - \frac{3}{8\pi^{2}} (C_{2}(A))^{2} + \frac{1}{\pi} \sum_{\Phi} \left[ \frac{S^{A}_{2}(\Phi)}{d_{\Phi A}} \left(X^{\Phi}_{a}\vartheta_{ab}X^{\Phi}_{b} + \right.\right.\\ \left.\left.  + \alpha_{B} \left(C^{B}_{2}(\Phi)+\frac{1}{2}\delta^{BA}C_{2}(A)\right)- \frac{1}{16\pi} \sum_{\Psi\Xi} (Y^{*\Phi\Psi\Xi}Y^{\Phi\Psi\Xi})\right)\right]
\end{array}\label{2lsA}\end{equation}

\begin{equation}\begin{array}{l}
(b_{d e})^{2-loop}_{SUSY}= \frac{1}{\pi} \sum_{\Phi}\left[X^{\Phi}_{d} X^{\Phi}_{e} \left(X^{\Phi}_{a}\vartheta_{ab}X^{\Phi}_{b} + \right.\right.\\ \left.\left. + \alpha_{B} C^{B}_{2}(\Phi)- \frac{1}{16\pi} \sum_{\Psi\Xi} (Y^{*\Phi\Psi\Xi}Y^{\Phi\Psi\Xi})\right)\right] \end{array}
\label{2lsde}\end{equation}
At 1-loop in SUSY case we have the following RGEs for Yukawa couplings:

\begin{equation}\begin{array}{l}
\mu\frac{d}{d \mu} Y^{\Phi\Psi\Xi} = \frac{1}{32\pi^{2}} \sum_{\Omega\Lambda\Delta} \left(Y^{\Phi\Psi\Omega}Y^{*\Omega\Lambda\Delta}Y^{\Xi\Lambda\Delta} + \right.\\ \left. + Y^{\Phi\Xi\Omega}Y^{*\Omega\Lambda\Delta}Y^{\Psi\Lambda\Delta} + Y^{\Xi\Psi\Omega}Y^{*\Omega\Lambda\Delta}Y^{\Phi\Lambda\Delta}\right) +\\ -\frac{1}{2\pi}Y^{\Phi\Psi\Xi} \left[\alpha_{A}\left(C^{A}_{2}(\Phi)+C^{A}_{2}(\Psi)+C^{A}_{2}(\Xi)\right) + \right.\\ \left. + \vartheta_{ab} \left(X^{\Phi}_{a}X^{\Phi}_{b}+X^{\Psi}_{a}X^{\Psi}_{b} +X^{\Xi}_{a}X^{\Xi}_{b}\right) \right]  
\end{array}\label{1lsY}\end{equation}
\newpage
\section{Tables of field multiplets}{\label{ApTab}}
In this section there are tables of field multiplets. Colors have been used to denote symmetry breakings. Broken gauge group and higgs fields that break it have the same color.
\begin{table*}[!htbp]
\caption{Multiplets $16$, $10$ and $\overline{126}$ under the $SO(10) \rightarrow SU(5) \oplus U(1)_{X}$ symmetry breaking.}
\begin{math}\begin{array}{|c|c|c|c||} 
\hline \hline \textcolor{violet}{SO(10)} & \textcolor{blue}{SU(5)} \oplus U(1)_{X} & SU(3)_{c} \oplus SU(2)_{L} \oplus  & SU(3)_{c} \oplus \textcolor{brown}{SU(2)_{L} \oplus U(1)_{Y}} \\ \textcolor{violet}{!}\footnote[1]{} & \textcolor{blue}{!} & \textcolor{red}{U(1)_{\widehat{Y}} \oplus U(1)_{X}} & \\  \hline
\hline             & \bar{5}\stackrel{}{\left(-\frac{3\sqrt{10}}{20}\right)}     & d_{R}^{*}\left[\bar{3},1\right]\left(\frac{\sqrt{15}}{15}\right)\left(-\frac{3\sqrt{10}}{20}\right)     & d_{R}^{*}\left[\bar{3},1\right]\left(\frac{1}{3}\right)\\ 
\cline{3-4}        &               & L\left[1,2\right]\stackrel{}{\left(-\frac{\sqrt{15}}{10}\right)}\left(-\frac{3\sqrt{10}}{20}\right)        & L\left[1,2\right]\left(-\frac{1}{2}\right)\\  
\cline{2-4}   16 &               & u_{R}^{*}\left[\bar{3},1\right]\left(-\frac{2\sqrt{15}}{15}\right)\stackrel{}{\left(\frac{\sqrt{10}}{20}\right)}     & u_{R}^{*}\left[\bar{3},1\right]\left(-\frac{2}{3}\right)\\  
\cline{3-4}        &   10\left(\frac{\sqrt{10}}{20}\right)        & Q\left[3,2\right]\left(\frac{\sqrt{15}}{30}\right)\stackrel{}{\left(\frac{\sqrt{10}}{20}\right)}         & Q\left[3,2\right]\left(\frac{1}{6}\right)\\ 
\cline{3-4}        &               & e_{R}^{*}\left[1,1\right]\left(\frac{\sqrt{15}}{5}\right)\stackrel{}{\left(\frac{\sqrt{10}}{20}\right)}     & e_{R}^{*}\left[1,1\right]\left(1\right)\\
\cline{2-4}       & 1\left(\frac{\sqrt{10}}{4}\right) & \nu_{R}^{*}\left[1,1\right]\left(0\right)\stackrel{}{\left(\frac{\sqrt{10}}{4}\right)}   & \nu_{R}^{*}\left[1,1\right]\left(0\right)\\  
\hline       &  & T_{u}\left[3,1\right]\stackrel{}{\left(-\frac{\sqrt{15}}{15}\right)}\left(-\frac{\sqrt{10}}{10}\right)\,!\footnote[1]{}   & \text{integrated out\footnote[2]{}}\\
&5_{u}\left(-\frac{\sqrt{10}}{10}\right)!&\text{integrated out\footnote[3]{}}&\\ 
\cline{3-4}   10! &               & H^{u}_{10}\left[1,2\right]\stackrel{}{\left(\frac{\sqrt{15}}{10}\right)}\left(-\frac{\sqrt{10}}{10}\right)     & \textcolor{brown}{H^{u}_{10}}\left[1,\textcolor{brown}{2}\right]\textcolor{brown}{\left(\frac{1}{2}\right)}\\  
\cline{2-4}        &  & T_{d}^{*}\left[\bar{3},1\right]\stackrel{}{\left(\frac{\sqrt{15}}{15}\right)}\left(\frac{\sqrt{10}}{10}\right)\,! & \text{integrated out}\\
&\bar{5}_{d}\left(\frac{\sqrt{10}}{10}\right)!&\text{integrated out}&\\  
\cline{3-4}        &               & H^{d}_{10}\left[1,2\right]\stackrel{}{\left(-\frac{\sqrt{15}}{10}\right)}\left(\frac{\sqrt{10}}{10}\right)     & \textcolor{brown}{H^{d}_{10}}\left[1,\textcolor{brown}{2}\right]\textcolor{brown}{\left(-\frac{1}{2}\right)} \\
\hline          & 1_{\overline{126}}\stackrel{}{\left(-\frac{\sqrt{10}}{2}\right)}       &  \textcolor{red}{\chi_{-}}\left[1,1\right] \textcolor{red}{\left(0\right)\stackrel{}{\left(-\frac{\sqrt{10}}{2}\right)}} & \chi_{-}\left[1,1\right]\left(0\right) = \textcolor{red}{h}\footnote[4]{} + \textcolor{red}{G}\footnote[4]{} \\  
\cline{2-4} & 5_{\overline{126}}\stackrel{}{\left(-\frac{\sqrt{10}}{10}\right)}! & \text{heavy combination of}\,5_{u}& \text{heavy combination of}\,5_{u} \\
&\text{mixing with}\,5_{u}&\text{and}\,5_{\overline{126}}\,\text{integrated out} & \text{and}\,5_{\overline{126}}\,\text{integrated out}  \\ 
\cline{2-4}  \overline{126}! &\overline{10}_{\overline{126}}\stackrel{}{\left(-\frac{3\sqrt{10}}{10}\right)} & & \\  
\cline{2-2}        & 15_{\overline{126}}\stackrel{}{\left(\frac{3\sqrt{10}}{10}\right)} &\text{integrated out} & \text{integrated out} \\ 
\cline{2-2}        & \overline{45}_{\overline{126}}\stackrel{}{\left(\frac{\sqrt{10}}{10}\right)}! &       &  \\ 
\cline{2-2}          & 50_{\overline{126}}\stackrel{}{\left(-\frac{\sqrt{10}}{10}\right)} & &   \\ 
\hline \hline  \end{array}\end{math}
\label{tab2}
\end{table*}
\footnotetext[1]{Mediators of proton decay are marked with "!".}\footnotetext[2]{When a field is assumed to have physical mass greater than the scale of a given symmetry breaking, the table doesn't include the representation of this field with respect to the residual gauge group that survive this particular symmetry breaking. Instead, it's written that the field is already "integrated out" in the whole scale-interval related to the residual group.}\footnotetext[3]{In the special case, when a field is assumed to have physical mass approximately equal to the scale of a given symmetry breaking, the table includes the representation of this field with respect to the residual gauge group that survive this particular symmetry breaking together with the "integrated out" annotation.}\footnotetext[4]{Symbol $h$ denotes a physical Higgs field and symbol $G$ denotes a Goldstone boson.}
\newpage
\begin{table*}[!htbp]
\caption{Multiplets $54$ and $210$ under the $SO(10) \rightarrow SU(5) \oplus U(1)_{X}$ symmetry breaking.}
\begin{math}\begin{array}{|c|c|c|c||} 
\hline \hline \textcolor{violet}{SO(10)} & \textcolor{blue}{SU(5)} \oplus U(1)_{X} & SU(3)_{c} \oplus SU(2)_{L} \oplus  & SU(3)_{c} \oplus \textcolor{brown}{SU(2)_{L} \oplus U(1)_{Y}} \\ \textcolor{violet}{!}& \textcolor{blue}{!} & \textcolor{red}{U(1)_{\widehat{Y}} \oplus U(1)_{X}} & \\  \hline
\hline        & & \left[1,1\right]\left(0\right)\left(0\right) \supset\textcolor{blue}{h}  &  \\
&&\text{integrated out}&\\
\cline{3-3}    & & \left[\bar{3},2\right]\stackrel{}{\left(\frac{\sqrt{15}}{6}\right)}\left(0\right) \supset\textcolor{blue}{G} & \\ 
&&\text{integrated out}&\\
\cline{3-3}    & \textcolor{blue}{24_{54}}\left(0\right) & \left[3,2\right]\stackrel{}{\left(-\frac{\sqrt{15}}{6}\right)}\left(0\right) \supset\textcolor{blue}{G} &  \text{integrated out} \\ 
&&\text{integrated out}&\\
\cline{3-3} 54 & & \left[1,3\right]\left(0\right)\left(0\right)  &  \\ 
&&\text{integrated out}&\\
\cline{3-3}    & & \left[8,1\right]\left(0\right)\left(0\right)  & \\ 
&&\text{integrated out}&\\ 
\cline{2-4}    &  15_{54}\stackrel{}{\left(-\frac{\sqrt{10}}{5}\right)}  & \text{integrated out} & \text{integrated out} \\ 
\cline{2-2}    &  \overline{15}_{54}\stackrel{}{\left(\frac{\sqrt{10}}{5}\right)} &  &  \\

\hline  & \textcolor{blue}{24_{210}}\left(0\right) & \text{heavy combination of}\,\textcolor{blue}{24_{54}} & \text{heavy combination of}\,\textcolor{blue}{24_{54}} \\ 
 & \text{mixing with}\,\textcolor{blue}{24_{54}} & \text{and}\,\textcolor{blue}{24_{210}}\,\text{integrated out}  & \text{and}\,\textcolor{blue}{24_{210}}\,\text{integrated out}  \\ 
\cline{2-4} & 5_{210}\stackrel{}{\left(\frac{2\sqrt{10}}{5}\right)}  & &  \\ 
\cline{2-2} \textcolor{violet}{210} & \overline{5}_{210}\stackrel{}{\left(-\frac{2\sqrt{10}}{5}\right)} & & \\
\cline{2-2} & 40_{210}\stackrel{}{\left(\frac{\sqrt{10}}{5}\right)} &  & \\ 
\cline{2-2} & \overline{40}_{210}\stackrel{}{\left(-\frac{\sqrt{10}}{5}\right)} &  &  \\
\cline{2-2} & 75_{210}\left(0\right) & \text{integrated out} & \text{integrated out} \\
\cline{2-2} & 1_{210}\left(0\right) \supset\textcolor{violet}{h}  &  &  \\ 
&\text{integrated out}&&\\
\cline{2-2} & 10_{210}\stackrel{}{\left(-\frac{\sqrt{10}}{5}\right)} \supset\textcolor{violet}{G}  &  &  \\
&\text{integrated out}&&\\
\cline{2-2} & \overline{10}_{210}\stackrel{}{\left(\frac{\sqrt{10}}{5}\right)} \supset\textcolor{violet}{G}  &  &  \\
&\text{integrated out}&&\\ 
\hline \hline  \end{array}\end{math}
\label{tab3}
\end{table*}
\newpage
\begin{table*}[!htbp]
\caption{Multiplets $16$ and $10$ under the $SO(10) \rightarrow SU(3)_{c} \oplus SU(2)_{L} \oplus  SU(2)_{R} \oplus U(1)_{\widehat{B-L}}$ symmetry breaking.}
\begin{math}\begin{array}{||c|c|c|c||} 
\hline \hline \textcolor{violet}{SO(10)} & SU(3)_{c} \oplus SU(2)_{L} \oplus  & SU(3)_{c} \oplus SU(2)_{L} \oplus & SU(3)_{c} \oplus \textcolor{brown}{SU(2)_{L} \oplus U(1)_{Y}}\\
\textcolor{violet}{!}& \textcolor{blue}{SU(2)_{R}} \oplus U(1)_{\widehat{B-L}} & \textcolor{red}{U(1)_{R} \oplus U(1)_{\widehat{B-L}}}  & \\  \hline
\hline             &   Q_{R}\left[\bar{3},1,2\right]\left(-\frac{\sqrt{6}}{12}\right)   & d_{R}^{*}\left[\bar{3},1\right]\left(\frac{1}{2}\right)\stackrel{}{\left(-\frac{\sqrt{6}}{12}\right)}     &  d_{R}^{*}\left[\bar{3},1\right]\left(\frac{1}{3}\right)\\ 
\cline{3-4}        &            &    u_{R}^{*}\left[\bar{3},1\right]\left(-\frac{1}{2}\right)\stackrel{}{\left(-\frac{\sqrt{6}}{12}\right)}      & u_{R}^{*}\left[\bar{3},1\right]\left(-\frac{2}{3}\right)\\  
\cline{2-4}   16 &       L\left[1,2,1\right]\stackrel{}{\left(-\frac{\sqrt{6}}{4}\right)}       & L\left[1,2\right]\left(0\right)\left(-\frac{\sqrt{6}}{4}\right)    & L\left[1,2\right]\left(-\frac{1}{2}\right)\\  
\cline{2-4}        &   Q\left[3,2,1\right]\stackrel{}{\left(\frac{\sqrt{6}}{12}\right)}        & Q\left[3,2\right]\left(0\right)\left(\frac{\sqrt{6}}{12}\right)         & Q\left[3,2\right]\left(\frac{1}{6}\right)\\  
\cline{2-4}        &       L_{R}\left[1,1,2\right]\left(\frac{\sqrt{6}}{4}\right)         & e_{R}^{*}\left[1,1\right]\left(\frac{1}{2}\right)\stackrel{}{\left(\frac{\sqrt{6}}{4}\right)}     & e_{R}^{*}\left[1,1\right]\left(1\right)\\
\cline{3-4}       &  & \nu_{R}^{*}\left[1,1\right]\stackrel{}{\left(-\frac{1}{2}\right)\left(\frac{\sqrt{6}}{4}\right)}   & \nu_{R}^{*}\left[1,1\right]\left(0\right)\\  
\hline             &  H_{10}\left[1,2,2\right]\left(0\right)   &  H^{d}_{10}\left[1,2\right]\stackrel{}{\left(-\frac{1}{2}\right)}\left(0\right)     & \textcolor{brown}{H^{d}_{10}}\left[1,\textcolor{brown}{2}\right]\textcolor{brown}{\left(-\frac{1}{2}\right)}  \\ 
\cline{3-4}   10! &               & H^{u}_{10}\left[1,2\right]\left(\frac{1}{2}\right)\left(0\right)     & \textcolor{brown}{H^{u}_{10}}\left[1,\textcolor{brown}{2}\right]\textcolor{brown}{\stackrel{}{\left(\frac{1}{2}\right)}}\\  
\cline{2-4}        & T_{u}\left[3,1,1\right]\stackrel{}{\left(-\frac{\sqrt{6}}{6}\right)}\,! &  &  \\
&\text{integrated out}&\text{integrated out}&\text{integrated out}\\  
\cline{2-2}        & T_{d}^{*}\left[\bar{3},1,1\right]\stackrel{}{\left(\frac{\sqrt{6}}{6}\right)}\,!    &    &  \\
&\text{integrated out}&&\\ 
\hline \hline  \end{array}\end{math}
\label{tab5}
\end{table*}
\newpage
\begin{table*}[!htbp]
\caption{Multiplet $\overline{126}$ under the $SO(10) \rightarrow SU(3)_{c} \oplus SU(2)_{L} \oplus  SU(2)_{R} \oplus U(1)_{\widehat{B-L}}$ symmetry breaking.}
\begin{math}\begin{array}{||c|c|c|c||} 
\hline \hline \textcolor{violet}{SO(10)} & SU(3)_{c} \oplus SU(2)_{L} \oplus  & SU(3)_{c} \oplus SU(2)_{L} \oplus & SU(3)_{c} \oplus \textcolor{brown}{SU(2)_{L} \oplus U(1)_{Y}}\\
\textcolor{violet}{!}& \textcolor{blue}{SU(2)_{R}} \oplus U(1)_{\widehat{B-L}} & \textcolor{red}{U(1)_{R} \oplus U(1)_{\widehat{B-L}}}  & \\  \hline
\hline & &\textcolor{red}{\chi_{-}}\left[1,1\right]\textcolor{red}{\left(1\right)\stackrel{}{\left(-\frac{\sqrt{6}}{2}\right)}}  & \chi_{-}\left[1,1\right]\left(0\right) = \textcolor{red}{h} + \textcolor{red}{G}\\  
\cline{3-4}        & T^{\chi}_{-}\left[1,1,3\right]\left(-\frac{\sqrt{6}}{2}\right) & \chi_{-}^{0}\left[1,1\right]\left(0\right)\stackrel{}{\left(-\frac{\sqrt{6}}{2}\right)} &   \\
&&\text{integrated out}&\text{integrated out}\\   
\cline{3-3}        &  & \chi_{-}^{-}\left[1,1\right]\left(-1\right)\stackrel{}{\left(-\frac{\sqrt{6}}{2}\right)} &  \\
&&\text{integrated out}&\\   
\cline{2-4} & H_{\overline{126}}\left[1,2,2\right]\left(0\right)&\text{heavy combination of}\,H_{10} &\text{heavy combination of}\,H_{10} \\
&\text{mixing with}\,H_{10} & \text{and}\,H_{\overline{126}}\,\text{integrated out} & \text{and}\,H_{\overline{126}}\,\text{integrated out} \\
\cline{2-4}        &  T^{L}_{+}\left[1,3,1\right]\stackrel{}{\left(\frac{\sqrt{6}}{2}\right)} &  &  \\ 
\cline{2-2}        & T^{d\,*}_{\overline{126}}\left[\bar{3},1,1\right]\stackrel{}{\left(\frac{\sqrt{6}}{6}\right)}\,! &  &  \\
&\text{integrated out}&&\\
\cline{2-2} \overline{126}! & T^{u}_{\overline{126}}\left[3,1,1\right]\stackrel{}{\left(-\frac{\sqrt{6}}{6}\right)}\,! &  &  \\
&\text{integrated out}&&\\  
\cline{2-2}        & \left[3,1,3\right]_{\overline{126}}\stackrel{}{\left(-\frac{\sqrt{6}}{6}\right)} &  &  \\
\cline{2-2}        & \left[6,1,3\right]_{\overline{126}}\stackrel{}{\left(\frac{\sqrt{6}}{6}\right)} & \text{integrated out} & \text{integrated out} \\
\cline{2-2}     & \left[\bar{3},3,1\right]_{\overline{126}}\stackrel{}{\left(\frac{\sqrt{6}}{6}\right)} &  &  \\
\cline{2-2}        & \left[\bar{6},3,1\right]_{\overline{126}}\stackrel{}{\left(-\frac{\sqrt{6}}{6}\right)} &  &  \\
\cline{2-2}        & \left[\bar{3},2,2\right]_{\overline{126}}\stackrel{}{\left(-\frac{\sqrt{6}}{3}\right)} &  &  \\
\cline{2-2}        & \left[3,2,2\right]_{\overline{126}}\stackrel{}{\left(\frac{\sqrt{6}}{3}\right)} &  &  \\
\cline{2-2}        & \left[8,2,2\right]_{\overline{126}}\left(0\right) &  &  \\    
\hline \hline  \end{array}\end{math}
\label{tab8}
\end{table*}
\newpage
\begin{table*}[!htbp]
\caption{Two $45$ multiplets under the $SO(10) \rightarrow SU(3)_{c} \oplus SU(2)_{L} \oplus  SU(2)_{R} \oplus U(1)_{\widehat{B-L}}$ symmetry breaking.}
\begin{math}\begin{array}{||c|c|c|c||} 
\hline \hline \textcolor{violet}{SO(10)} & SU(3)_{c} \oplus SU(2)_{L} \oplus  & SU(3)_{c} \oplus SU(2)_{L} \oplus & SU(3)_{c} \oplus \textcolor{brown}{SU(2)_{L} \oplus U(1)_{Y}}\\
\textcolor{violet}{!} & \textcolor{blue}{SU(2)_{R}} \oplus U(1)_{\widehat{B-L}} & \textcolor{red}{U(1)_{R} \oplus U(1)_{\widehat{B-L}}}  & \\
\hline \hline   &   & \left[1,1\right]\left(0\right)\left(0\right)\supset\textcolor{blue}{h}  &  \\ 
&&\text{integrated out}&\\ 
\cline{3-3}& \textcolor{blue}{T^{A}_{45}}\left[1,1,\textcolor{blue}{3}\right]\left(0\right)&\left[1,1\right]\left(1\right)\left(0\right)\supset\textcolor{blue}{G} & \text{integrated out} \\
&&\text{integrated out}&\\ 
\cline{3-3}      &   & \left[1,1\right]\left(-1\right)\left(0\right)\supset\textcolor{blue}{G} &  \\ 
&&\text{integrated out}&\\ 
\cline{2-4}        & \left[1,3,1\right]\left(0\right) &  &  \\
\cline{2-2}        & \left[8,1,1\right]\left(0\right) &  &  \\
\cline{2-2}        & \left[1,1,1\right]\left(0\right) &  &  \\
\cline{2-2}  45    & \left[\bar{3},1,1\right]\stackrel{}{\left(-\frac{\sqrt{6}}{3}\right)} &  &  \\
\cline{2-2}      & \left[3,1,1\right]\stackrel{}{\left(\frac{\sqrt{6}}{3}\right)} & \text{integrated out} & \text{integrated out} \\
\cline{2-2}      & \left[\bar{3},2,2\right]\stackrel{}{\left(\frac{\sqrt{6}}{6}\right)} &  &  \\  
\cline{2-2}      & \left[3,2,2\right]\stackrel{}{\left(-\frac{\sqrt{6}}{6}\right)} &  &  \\
\hline    & \textcolor{blue}{T^{B}_{45}} \left[1,1,\textcolor{blue}{3}\right]\left(0\right)& \text{heavy combination of}\,\textcolor{blue}{T^{A}_{45}} & \text{heavy combination of}\,\textcolor{blue}{T^{A}_{45}} \\ 
    &  \text{mixing with}\,\textcolor{blue}{T^{A}_{45}} & \text{and}\,\textcolor{blue}{T^{B}_{45}}\,\text{integrated out} & \text{and}\,\textcolor{blue}{T^{B}_{45}}\,\text{integrated out} \\
\cline{2-4}        & \left[1,3,1\right]\left(0\right) &  &  \\
\cline{2-2}        & \left[8,1,1\right]\left(0\right) &  &  \\
\cline{2-2}        & \left[1,1,1\right]\left(0\right) \supset\textcolor{violet}{h}  &  &  \\
&\text{integrated out}&&\\ 
\cline{2-2}\textcolor{violet}{45}& \left[\bar{3},1,1\right]\stackrel{}{\left(-\frac{\sqrt{6}}{3}\right)}\supset\textcolor{violet}{G} &  &\\
&\text{integrated out}&&\\ 
\cline{2-2}   & \left[3,1,1\right]\stackrel{}{\left(\frac{\sqrt{6}}{3}\right)}\supset\textcolor{violet}{G} & \text{integrated out} & \text{integrated out} \\
&\text{integrated out}&&\\ 
\cline{2-2}        & \left[\bar{3},2,2\right]\stackrel{}{\left(\frac{\sqrt{6}}{6}\right)}\supset\textcolor{violet}{G} &  &  \\
&\text{integrated out}&&\\   
\cline{2-2}        & \left[3,2,2\right]\stackrel{}{\left(-\frac{\sqrt{6}}{6}\right)}\supset\textcolor{violet}{G} &  &  \\
&\text{integrated out}&&\\     
\hline \hline \end{array}\end{math}
\label{tab6}
\end{table*}
\newpage
\section{Superpotentials and mass terms}{\label{ApW}}

The most general form of the superpotential in Case I is the following
\begin{equation}\begin{array}{l}
W_{I}=\frac{1}{2}\,16_{i}\,(Y_{10}^{ij}10+Y_{\overline{126}}^{ij}\overline{126})\,16_{j} + \\
+ Y_{A} (210\,210\,210) +  Y_{B} (210\,210\,54) +  Y_{C} (210\,126\,\overline{126}) + \\+ Y_{D} (210\,10\,126) +  Y_{\overline{D}} (210\,10\,\overline{126}) 
+ Y_{E} (54\,54\,54) +\\+ Y_{F} (54\,126\,126) + Y_{\overline{F}} (54\,\overline{126}\,\overline{126}) + Y_{G} (54\,10\,10)+\\
+ M_{210}(210\,210) + M_{54}(54\,54) + M_{126}(126\,\overline{126}) +\\+ M_{10}(10\,10)
\end{array}\end{equation}
The most general form of the superpotential in Case II is written below
\begin{equation}\begin{array}{l}
W_{II}=\frac{1}{2}\,16_{i}\,(Y_{10}^{ij}10+Y_{\overline{126}}^{ij}\overline{126})\,16_{j} + \\
+ Y_{K}^{\alpha\beta\gamma} (45_{\alpha}\,45_{\beta}\,45_{\gamma}) + Y_{L}^{\alpha} (45_{\alpha}\,126\,\overline{126}) + Y_{M}^{\alpha} (45_{\alpha}\,10\,10)+ \\
+ M_{45}^{\alpha\beta}(45_{\alpha}\,45_{\beta}) + M_{126}(126\,\overline{126}) + M_{10}(10\,10)
\end{array}\end{equation}
$i,j \in \left\{1,2,3\right\}$ and $\alpha,\beta,\gamma \in \left\{1,2\right\}$

Each mass parameter proportional to $\mu_{0}$ is generated by the $SO(10)$-breaking VEV of the intermediate group singlet which is $1_{210}$ in Case I and $[1,1,1]_{45_{1}}$ in Case II. Parts of superpotentials that contain these singlets are the following
\begin{equation}\begin{array}{l}
W_{I}\supset 1_{210} \left[Y_{A} \sum_{X\subset 210}m_{X}(X_{210}\,\overline{X}_{210}) +  2Y_{B} (24_{54}\,24_{210})+\right.\\
 \left.+  Y_{C} \sum_{X\subset 126}(X_{126}\,\overline{X}_{\overline{126}}) +  Y_{D} (5_{10}\,\overline{5}_{126}) +  Y_{\overline{D}} (\overline{5}_{10}\,5_{\overline{126}})\right]
\end{array}\end{equation}

\begin{equation}\begin{array}{l}
W_{II} \supset [1,1,1]_{45_{1}} \left[Y_{K}^{1\beta\gamma} \sum_{X\subset 45}n^{\beta\gamma}_{X}(X_{45_{\beta}}\,\overline{X}_{45_{\gamma}}) +\right.\\
 \left.+ Y_{L}^{1} \sum_{X\subset 126}(X_{126}\,\overline{X}_{\overline{126}}) + Y_{M}^{1}  \sum_{X\subset 10}(X_{10}\,\overline{X}_{10}) \right]
\end{array}\end{equation}
$m_{X}$ and $n^{\beta\gamma}$ are appropriate combinatorial factors. The only fields that don't couple to the intermediate group singlet are $15_{54}$ and $\overline{15}_{54}$
in Case I.

\section{CMS limits}{\label{ApCMS}}

In this section we show in detail how to transform experimental limits given by CMS \cite{CMS} into limits on $g'_{B-L}(M_{Z'})$ and $g_{B-L}(M_{Z'})$. In \cite{CMS} we can see upper limit on the quantity denoted by $R_{\sigma}$ and its dependence on $M_{Z'}$. $R_{\sigma}$ is defined by the following formula
\begin{equation}
R_{\sigma} = \frac{\sigma(pp\rightarrow Z' + X \rightarrow l\bar{l} + X)}{\sigma(pp\rightarrow Z + X \rightarrow l\bar{l} + X)}
\label{CC1}\end{equation}
$X$ is anything and $l$ is a an electron or a muon. The limit on $R_{\sigma}$ comes from data collected in $7$ TeV and $8$ TeV runs. It's explicitly written that the value of $\sqrt{s}$, that should be inserted to $\sigma(pp\rightarrow Z' + X \rightarrow l\bar{l} + X)$ and $\sigma(pp\rightarrow Z + X \rightarrow l\bar{l} + X)$, is equal to $8$ TeV.
In the narrow width approximation the formula for $R_{\sigma}$ simplifies to
\begin{equation}
R_{\sigma} = \frac{\sigma(pp\rightarrow Z' + X)\cdot BR(Z'\rightarrow l\bar{l})}{\sigma(pp\rightarrow Z + X)\cdot BR(Z\rightarrow l\bar{l})}
\label{CC2}\end{equation}
We calculated $\sigma(pp\rightarrow Z' + X)$, $BR(Z'\rightarrow l\bar{l})$, $\sigma(pp\rightarrow Z + X)$ and $BR(Z\rightarrow l\bar{l})$ in the leading order (LO) of the perturbation expansion as functions of $g'_{B-L}(M_{Z'})$ and $g_{B-L}(M_{Z'})$. To do it properly, we took into account that $g'_{B-L}$ and $g_{B-L}$ are coupling constants of the $Z'^{0}$ boson (\ref{L2low}) and not the physical $Z'$ boson. The most convenient way to describe couplings of the physical $Z'$ boson to fermions (and scalars) is the following
\begin{equation}\begin{array}{l}
L_{Z'\,int} \supset Z'^{\mu} \sum_{f} \bar{\Psi}_{f}\gamma_{\mu} \\ \left((B-L)^{f} g^{Z'}_{B-L} + Q^{f} g^{Z'}_{Q} + T_{3}^{f} P_{L} g^{Z'}_{T_{3}}\right)\Psi_{f}
\label{CC3}\end{array}\end{equation}
with three fermion-independent coupling constants -\newline $g^{Z'}_{B-L}$, $g^{Z'}_{Q}$ and $g^{Z'}_{T_{3}}$. They can be expressed in the following way
\begin{equation}\begin{array}{l}
\textcolor{white}{\stackrel{I}{I}} g^{Z'}_{B-L} = \cos\theta'\, g_{B-L} \\
\textcolor{white}{\stackrel{I}{I}} g^{Z'}_{Q} = \cos\theta'\, g'_{B-L} + \sin^{2}\theta_{W}\, \sin\theta'\, g_{Z} \\
\textcolor{white}{\stackrel{I}{I}} g^{Z'}_{T_{3}} = -\cos\theta'\, g'_{B-L} - \sin\theta'\, g_{Z} 
\end{array}\end{equation}
$\theta'$ is the $Z-Z'$ mixing angle. It is set by the diagonalization of the mass matrix for initial $Z^{0}$ and $Z'^{0}$  bosons. Physical $Z$ and $Z'$ bosons are eigenstates of this matrix. $\theta'$ can be expressed as a function of $g'_{B-L}$ and $M_{Z'}$.

\begin{equation}\begin{array}{l}
\cos 2\theta' = \frac{g'^{2}_{B-L}(M_{Z'}^{2}+M_{Z}^{2})+g_{Z}\sqrt{g_{Z}^{2}(M_{Z'}^{2}-M_{Z}^{2})^{2}-4g'^{2}_{B-L}M_{Z'}^{2}M_{Z}^{2}}}{(g_{Z}^{2}+g'^{2}_{B-L})(M_{Z'}^{2}-M_{Z}^{2})} \\
\textcolor{white}{.}\\
\sin 2\theta' = \frac{g'_{B-L}g_{Z}(M_{Z'}^{2}+M_{Z}^{2})-g'_{B-L}\sqrt{g_{Z}^{2}(M_{Z'}^{2}-M_{Z}^{2})^{2}-4g'^{2}_{B-L}M_{Z'}^{2}M_{Z}^{2}}}{(g_{Z}^{2}+g'^{2}_{B-L})(M_{Z'}^{2}-M_{Z}^{2})}
\end{array}\end{equation}

Calculating $\sigma(pp\rightarrow Z + X)$ and $BR(Z\rightarrow l\bar{l})$ we took into account effects of the $Z-Z'$ mixing on the $Z$ boson couplings
\begin{equation}\begin{array}{l}
\textcolor{white}{\stackrel{I}{I}} g^{Z}_{B-L} = \sin\theta'\, g_{B-L} \\
\textcolor{white}{\stackrel{I}{I}} g^{Z}_{Q} = \sin\theta'\, g'_{B-L} - \sin^{2}\theta_{W}\, \cos\theta'\, g_{Z} \\
\textcolor{white}{\stackrel{I}{I}} g^{Z}_{T_{3}} = -\sin\theta'\, g'_{B-L} + \cos\theta'\, g_{Z} 
\end{array}\end{equation}
To obtain $BR(Z'\rightarrow l\bar{l})$ at LO we calculated all needed decay widths: $\Gamma(Z'\rightarrow W^{+}W^{-})$, $\Gamma(Z'\rightarrow Z h)$ and $\Gamma(Z'\rightarrow f\bar{f})$ where $f$ is a SM-fermion. We assumed that masses of all three RH-neutrinos are larger than $M_{Z'}/2$ so we didn't include $\Gamma(Z'\rightarrow \nu_{R}\nu_{R})$. Masses of all fermions except for the $t$ quark have been neglected in these decay widths. We took $M_{h}=125.7$ GeV which is in good agreement with \cite{ATLASh} and \cite{CMSh}. For $BR(Z\rightarrow l\bar{l})$ the number of decay widths is much smaller: $\Gamma(Z \rightarrow f\bar{f})$ where $f$ is every SM-fermion except for the $t$ quark. For these decay widths we neglected masses of all fermions except for $b$ and $c$ quarks and $\tau$.

To  calculate $\sigma(pp\rightarrow Z' + X)$ and $\sigma(pp\rightarrow Z + X)$ we've used the MSTW 2008 PDF set \cite{PDFs} with the factorization scale equal to $M_{Z'}$ and $M_{Z}$ respectively. At LO all partonic contributions to these cross-sections have the form $\sigma(q\bar{q}\rightarrow Z')$ and $\sigma(q\bar{q}\rightarrow Z)$. $\sigma(t\bar{t}\rightarrow Z)$ is kinematicaly impossible and for $\sigma(t\bar{t}\rightarrow Z')$ the corresponding PDF is equal to $0$. All other quarks have been taken into account and their masses have been neglected with two exceptions - $\sigma(b\bar{b}\rightarrow Z)$ and $\sigma(c\bar{c}\rightarrow Z)$.


\begin{thebibliography}{}
\bibitem{Aguila} F. del Aguila, G.D. Coughlan, M. Quiros,
Nucl.Phys. \textbf{B307},(1988) 633, Erratum-ibid. \textbf{B312}, (1989) 751.
[\href{http://www.sciencedirect.com/science/article/pii/0550321388902660}{DOI: 10.1016/0550-3213(88)90266-0}]
\bibitem{Aguila2} F. del Aguila, J.M. Moreno, M. Quiros,
Nucl.Phys.Proc.Suppl. \textbf{16}, (1989) 621-623. [\href{http://www.sciencedirect.com/science/article/pii/0920563290906185}{DOI: 10.1016/0920-5632(90)90618-5}]
\bibitem{Leike} A. Leike, S. Riemann,
Nucl.Phys.Proc.Suppl. \textbf{29A}, (1992) 270-274. \newline [\href{http://www.sciencedirect.com/science/article/pii/092056329290453Y}{DOI: 10.1016/0920-5632(92)90453-Y}] 
\bibitem{Perez} F. del Aguila, M. Masip, M. Perez-Victoria,
Nucl.Phys. \textbf{B456}, (1995) 531-549. [\href{http://arxiv.org/pdf/hep-ph/9507455.pdf}{hep-ph/9507455}] 
\bibitem{Perez2} F. del Aguila, M. Masip, M. Perez-Victoria,
Acta Phys.Polon. \textbf{B27}, (1996) 1469-1478. [\href{http://arxiv.org/pdf/hep-ph/9603347.pdf}{hep-ph/9603347}]
\bibitem{Appelquist} Thomas Appelquist, Bogdan A. Dobrescu, Adam R. Hopper,
Phys.Rev. \textbf{D68}, (2002) 035012. [\href{http://arxiv.org/pdf/hep-ph/0212073v3.pdf}{hep-ph/0212073}]

\bibitem{Lopez} Jorge L. Lopez,
Nucl.Phys.Proc.Suppl. \textbf{52A}, (1996) 284-288. [\href{http://arxiv.org/pdf/hep-ph/9607231.pdf}{hep-ph/9607231}]
\bibitem{Zwirner1} Ennio Salvioni, Giovanni Villadoro, Fabio Zwirner,
JHEP \textbf{0911}, (2009) 068. \newline [\href{http://arxiv.org/pdf/0909.1320.pdf}{arXiv:0909.1320}]
\bibitem{Zwirner2} Ennio Salvioni, Alessandro Strumia, Giovanni Villadoro, Fabio Zwirner,
JHEP \textbf{1003}, (2010) 010. [\href{http://arxiv.org/pdf/0911.1450.pdf}{arXiv:0911.1450}]
\bibitem{Erler} Jens Erler, Paul Langacker, Shoaib Munir, Eduardo Rojas,
C11-04-11.2, (2011) [\href{http://arxiv.org/pdf/1108.0685.pdf}{arXiv:1108.0685}]
\bibitem{Kim} Jihn E. Kim, Seodong Shin,
Phys.Rev. \textbf{D85}, (2012) 015012. [\href{http://arxiv.org/pdf/1104.5500.pdf}{arXiv:1104.5500}]

\bibitem{slansky} R. Slansky,
Phys.Rept. \textbf{79}, (1981) 1-128. [\href{http://ccdb5fs.kek.jp/cgi-bin/img/allpdf?198102061}{KEK 198102061}]

\bibitem{Joshipura} Anjan S. Joshipura, Ketan M. Patel
Phys.Rev. \textbf{D83}, (2011) 095002. [\href{http://arxiv.org/pdf/1102.5148.pdf}{arXiv:1102.5148}]
\bibitem{Bertolini} Stefano Bertolini, Luca Di Luzio, Michal Malinsky,
Phys.Rev. \textbf{D80}, (2009) 015013. [\href{http://arxiv.org/pdf/0903.4049.pdf}{arXiv:0903.4049}]

\bibitem{Senjanovic} Goran Senjanovic
AIP Conf.Proc. \textbf{1200}, (2009) 131-141. [\href{http://arxiv.org/pdf/0912.5375.pdf}{arXiv:0912.5375}]
\bibitem{Jones} Jeff L. Jones,
Phys.Rev. \textbf{D79}, (2009) 075009. [\href{http://arxiv.org/pdf/0812.2106.pdf}{arXiv:0812.2106}]

\bibitem{Pokorski} Marcela S. Carena, S. Pokorski, C.E.M. Wagner,
Nucl.Phys. \textbf{B406}, (1993) 59-89. [\href{http://arxiv.org/pdf/hep-ph/9303202v1.pdf}{hep-ph/9303202}]
\bibitem{PDG} J. Beringer et al. (Particle Data Group), Phys.Rev. \textbf{D86}, (2012) 010001. [\href{http://pdg.lbl.gov/}{PDG}] [\href{http://pdg.lbl.gov/2012/reviews/rpp2012-rev-guts.pdf}{PDG-GUT}]
\bibitem{ATLAS} ATLAS Collaboration, 
ATLAS-CONF-2012-129, (2012) [\href{https://cdsweb.cern.ch/record/1477926/files/ATLAS-CONF-2012-129.pdf}{ATLAS $Z'$}]
\bibitem{CMS} CMS Collaboration,
CMS PAS EXO-12-015, (2012) [\href{http://cdsweb.cern.ch/record/1461216/files/EXO-12-015-pas.pdf}{CMS $Z'$}]
\bibitem{ATLASh} ATLAS Collaboration (Georges Aad et al.),
Phys.Lett. \textbf{B716}, (2012) 1-29. \newline 
[\href{http://arxiv.org/pdf/1207.7214.pdf}{arXiv:1207.7214}]
\bibitem{CMSh} CMS Collaboration (Serguei Chatrchyan et al.),
Phys.Lett. \textbf{B716}, (2012) 30-61. 
[\href{http://arxiv.org/pdf/1207.7235.pdf}{e-arXiv:1207.7235}]

\bibitem{PDFs} A.D. Martin, W.J. Stirling, R.S. Thorne, G. Watt, 
Eur.Phys.J. \textbf{C63}, (2009) 189-285.
[\href{http://arxiv.org/pdf/0901.0002.pdf}{arXiv:0901.0002}]
\bibitem{M_Z masses} Zhi-zhong Xing, He Zhang, Shun Zhou,
Phys.Rev. \textbf{D77}, (2008) 113016. \newline [\href{http://arxiv.org/pdf/0712.1419.pdf}{arXiv:0712.1419}]
\end{thebibliography}
\end {document}